\documentclass[fleqn,usenatbib]{mnras}

\usepackage{newtxtext,newtxmath}
\usepackage{multirow}
\usepackage{orcidlink}

\usepackage[T1]{fontenc}

\usepackage{color}   
\usepackage{ulem}
\definecolor{dgreen}{rgb}{0,0.5,0}
\definecolor{dp}{rgb}{0.5,0,0.5}
\definecolor{magen}{rgb}{0.79,0.08,0.48}
\definecolor{darkred}{rgb}{0.65,0.06,0.37}
\definecolor{brightred}{rgb}{1.0, 0.0, 0.0}
\usepackage[dvipsnames]{xcolor}

\usepackage{gensymb}
\usepackage{pdflscape}

\usepackage{orcidlink} 
\newcommand{\orcid}[1]{\href{https://orcid.org/#1}{\textcolor[HTML]{A6CE39}{\aiOrcid}}}

\DeclareRobustCommand{\VAN}[3]{#2}
\let\VANthebibliography\thebibliography
\def\thebibliography{\DeclareRobustCommand{\VAN}[3]{##3}\VANthebibliography}

\usepackage{graphicx}	
\usepackage{amsmath}	

\title[Modelling HS 0810+2554 with Multipoles]{A Fuzzy Situation Eased: Cold Dark Matter with Multipoles Can Explain The Double Radio Quad Lens HS 0810+2554}

\author[J. H. Miller Jr and L. L. R. Williams]{
John H. Miller Jr$^{1}$\thanks{E-mail: mill9614@umn.edu (JHMJ); llrw@umn.edu (LLRW)} \orcidlink{0000-0002-2901-5260}
and Liliya L. R. Williams$^{1}$ \orcidlink{0000-0002-6039-8706}
\\
$^{1}$Minnesota Institute for Astrophysics, University of Minnesota, 116 Church Street, Minneapolis, MN 55455, USA }

\date{Accepted XXX. Received YYY; in original form ZZZ}

\pubyear{2025}

\begin{document}
\label{firstpage}
\pagerange{\pageref{firstpage}--\pageref{lastpage}}
\maketitle

\begin{abstract} 
Originally observed in isophotal density contours of elliptical galaxies, higher order perturbations in the form of Fourier modes, or multipoles, are becoming increasingly recognized as necessary to account for angular mass complexity in strong lensing analyses. When smooth, elliptical CDM mass models fail, multipoles often emerge as solutions. With the discovery of two radio jets in the source quasar, the strong gravitational lens HS 0810+2554 can no longer be well fit by elliptical mass models, suggesting perturbations on small-scales. In this paper, we investigate the efficacy of multipoles $m=1$ (lopsidedness), $m=3$ (triangleness), and $m=4$ (boxiness and diskiness) in addressing the image positional anomalies of the two radio quads of HS 0810+2554. Due to the exact pairing and arrival sequence of the images being unknown, we consider all feasible image configurations. With 64 unique best-fit models, we achieve a fit of $\chi=1.59$ ($\chi^2=2.53$), with $m=1,3,4$ multipole strengths of 0.9\%, 0.4\%, and 0.6\%, respectively, with images in the reverse time ordering. Elliptical+shear models from previous works find $\chi\!\sim\!7\!-\!10$, for comparison. With the morphological (i.e., standard) arrival sequence, we achieve a fit of $\chi=2.95$ with two images being assigned to opposite sources. Therefore, CDM mass models with mass complexity in the form of multipoles are able to adequately explain the positional anomalies in HS 0810+2554. Alternative dark matter theories, like fuzzy dark matter, need not be invoked.

\end{abstract}


\section{Introduction} \label{sec:intro}

The standard cold dark matter (CDM) cosmological model has been successful in describing the structure of the Universe down to kpc scales, and the cosmic microwave background (CMB) power-spectra \citep{2020plancki}. On sub-galactic scales, however, N-body simulations of CDM have been discrepant with observations \citep[][for review]{2017bullock,2017popolo}. Various baryonic processes have been proposed as potential solutions to these failures, i.e., suppression of star formation at re-ionization \citep[e.g.,][]{2024mcquinn} and supernovae feedback \citep[e.g.,][]{2012pontzen}. Likewise, other dark matter frameworks have been suggested to resolve the various small-scale inconsistencies of CDM.

One potential solution to the `missing satellite' problem is warm dark matter (WDM), where the free-streaming length of dark matter is non-negligible, leading to the suppression of halos at low-masses. Observations of quadruply lensed quasars, or quads, have been used to constrain this free-streaming length and probe the subhalo mass fraction \citep{2019gilman,2020gilman,2020hseuh,2024keeley}. Additionally, quads have been utilized to constrain the interaction cross-section within the self-interacting dark matter (SIDM) framework \citep{2021gilman,2023gilman}. 
Collisionless CDM N-body simulations predict low-mass dark matter halos that are too dense and too cuspy, in possible contrast with theoretical expectations \citep{2025williams}.
Within simulations, SIDM can resolve this core-cusp problem by including dark matter scattering and creating cored central density profiles through collisional heat \citep{2005ahn,2015elbert}. However, neither WDM nor SIDM are prefect solutions to the small-scale problems of CDM.

An attractive prospective solution to both of these problems comes in the form of an ultralight bosonic particle of mass $10^{-22}$ eV, called fuzzy dark matter (FDM, or wave dark matter). Due to the low-mass nature of these particles they can be treated as classical waves on galactic scales with a de Broglie wavelength of $\lambda_\textup{dB} \approx$ 1 kpc. \citealt{2020schutz} has shown that the subhalo mass function predicts fewer low-mass halos than CDM, similar to that of WDM. Additionally, the central density of FDM halos are well described as a soliton with a core scale equal to the de Broglie wavelength \citep{2021li}. However, local ultra-faint dwarf galaxies cannot currently be explained with FDM \citep{Marsh2019,Dalal2022,2025benito}. Perturbations smaller than the de Broglie wavelength can occur, which could resolve flux anomalies \citep{2020chan} and positional anomalies in certain quad lenses. \citealt{2023amruth} used these perturbations to try to resolve the position anomalies in the galaxy lens HS 0810+2554.

\begin{figure}
    \centering
    \includegraphics[width=0.9\linewidth]{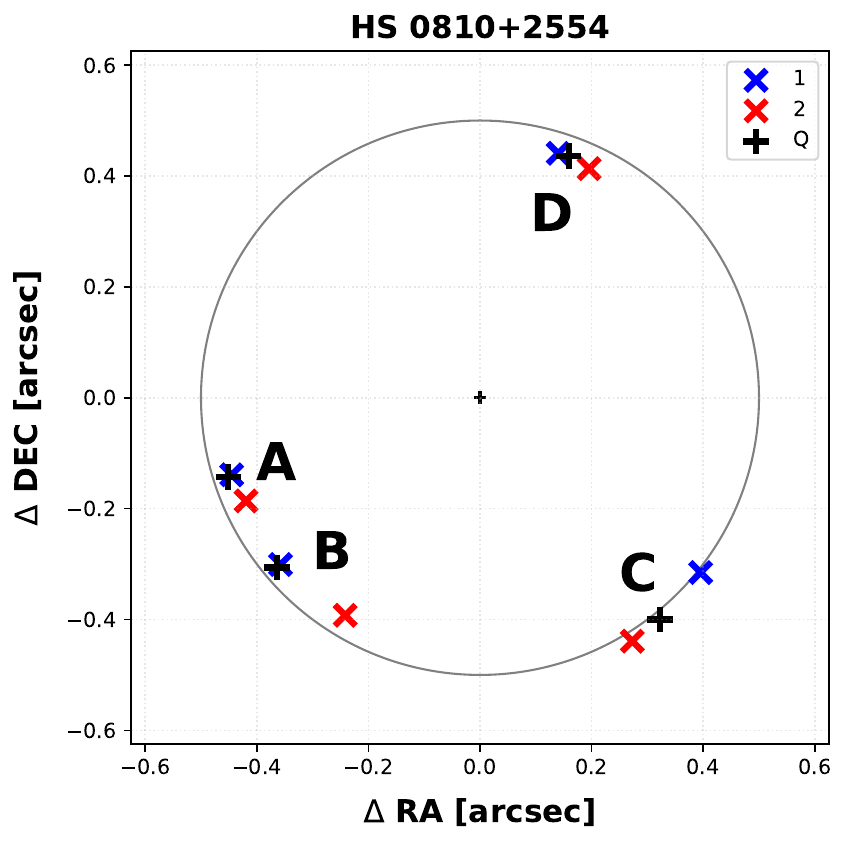}
    \caption{The gravitational lens HS 0810+2554. The image positions for the two quadruply lensed radio jets (blue and red `X's) and the quadruply lensed RQQ (large black `+'s) are reported in Table \ref{tab:positions}. A-D designates the relative brightness labelling scheme of the images. The images are centered on the CASTLES survey reported optical galaxy center \citep[small black `+';][]{1998kochanek}, assuming a nominal (+6 mas, +4 mas) offset for the radio images so that images A1 and AQ are not coincident. The gray circle has a radius of 0.5 arcsec and is included to show the near-circular nature of the system.}
    \label{fig:hs0810}
\end{figure}

The lens HS 0810+2554 (Figure \ref{fig:hs0810}) is a triple fold-configuration quad lens consisting of one optical quad discovered by \cite{2002reimer} with HST, and two radio quads discovered nearly two decades later by \cite{2019hartley}. The three sources are a radio quiet quasar and its two jets at a redshift of $z_s=1.51$. Numerous studies have been able to successfully reproduce the HST quad lens with a singular isothermal ellipse plus external shear \citep{2011assef,2014chartas,2015jackson,2016chartas,2020chartas,2020nierenberg}. However, no smooth mass model has been able to reconstruct the image positions of the two lensed radio jets, indicating that the galaxy lens is likely more complex than initially thought. It has been shown that these position anomalies could be explained by small-scale FDM perturbations \citep{2023amruth}. As for the CDM framework, no thorough analysis yet exists.

The standard strong lensing analysis under the CDM paradigm is to model the combined mass contribution of baryons and dark matter as a singular isothermal ellipse (SIE), or an elliptical power law (EPL), with external shear. Some papers separate the two mass components, modelling the baryons as a Sersic distribution \citep{2019chen}. However, these simply parameterized mass models neglect higher order perturbations often observed in the light distributions of lensing galaxies, which are primarily massive ellipticals. The stellar isophotes of massive elliptical galaxies are observed to have azimuthal perturbations beyond ellipticity in the form of multipoles \citep{2006hao,2009kormendy,2017mitsuda,2024amvrosiadis}, which can significantly differ in strength between the inner and outer regions of the galaxy \citep{2014chaware}. The most common of these multipoles is the fourth-order $m=4$, called `boxiness' and `diskiness', and has been reproduced in N-body simulations of galaxy mergers \citep{2005khochfar}. However, it is currently not well quantified to what degree these Fourier perturbations exist within the dark matter distribution itself. 

Recent strong lensing studies have highlighted the importance of including third- and fourth-order multipoles in mass model fitting. Both ALMA \citep{2024stacey} and VLBI \citep{2022powell} observations of extended source lenses found strong evidence for elliptical power-law models with $m=3,4$ multipoles when compared to ones without. The multipole $m=1$, or lopsidedness, which can be modelled as two offset mass components \citep{2018gomer,2024miller}, was required to achieve a $\chi^2\le1$ model for the quadruply lensed quasar WFI 2033-4723 \citep{2021barrera}. When not taken into account, multipoles have also been shown to bias detections of dark matter subhalos \citep{2024cohen,2024oriordan} and time-delay cosmography \citep{2022vyvere1,2022powell}. Additionally, mass models with insufficient internal complexity in the form of boxiness and diskiness can be incorrectly compensated for by external shear \citep{2024etherington}. When everything is considered, multipoles are likely necessary mass complexities when simple mass models fail.

In this paper, we investigate whether CDM mass models with Fourier perturbations in the form of multipoles $m=1,3,4$ can rectify the positional anomalies seen in previous smooth lens models of the eight radio images of HS 0810+2554. We create four main mass models with increasing complexity to explore various parameter combinations. First, we fit the eight images with an EPL and add a constant external shear to account for the contribution of a nearby galaxy group. All four models include this constant external shear. Next, we model the baryons and dark matter separately and include lopsidedness with the multipole $m=1$. Then, we create a model with contributions from all three multipoles $m=1,3,4$. Finally, the previous model is adjusted slightly by incorporating a free external shear component. Because the eight radio images are not source identified, we model all 16 possible image configurations. In total, we find 64 best-fit models. 

Observations of HS 0810+2554 and the lensing observables are discussed in Section \ref{sec:lensobs}. Section \ref{sec:mods} contains the lens model history of HS 0810 (Section \ref{sec:existing}) and the four main mass models used in our analysis (Section \ref{sec:ours}). The results of our model fitting are presented in Section \ref{sec:results}, with a discussion of our findings in Section \ref{sec:discussion}.

\section{Observables} \label{sec:lensobs}

Initially dubbed ``a bright twin'' to PG 1115+080, the gravitational lens HS 0810+2554 was discovered serendipitously with the Hubble Space Telescope (HST) and is a fold configuration lens with a singular optical quad \citep{2002reimer}. Nearly two decades later, two radio quads were revealed to be hiding in the nano-Jansky regime \citep{2019hartley}. In total, HS 0810+2554 (shown in Figure \ref{fig:hs0810}) contains three quad lenses, one in optical (black) and two in radio (blue and red), all originating from the same source redshift of $z_s$ = 1.51 \citep{2014chartas}. Numerous studies have been conducted to probe the source galaxy, which hosts a radio quiet quasar (RQQ) \citep{2011assef,2014chartas,2015jackson,2016chartas,2019hartley,2020chartas,2021tozzi}. However, not much is known about the lensing galaxy itself, likely attributed to being at an estimated redshift of $z_l$ = 0.89 \citep{2011mosquera}.

\begin{table}
    \centering
    \begin{tabular}{ccc} \hline
        $\theta_i$ & $\Delta$ RA [mas] & $\Delta$ DEC [mas] \\ \hline
        A1 & 0.0 $\pm$ 1.4 & 0.0 $\pm$ 1.0 \\
        A2 & 25.8 $\pm$ 1.2 & -47.2 $\pm$ 1.5 \\
        B1 & 87.5 $\pm$ 1.2 & -161.8 $\pm$ 0.8 \\
        B2 & 203.4 $\pm$ 1.0 & -253.4 $\pm$ 1.0 \\
        C1 & 840.5 $\pm$ 1.6 & -176.3 $\pm$ 1.2 \\
        C2 & 717.9 $\pm$ 1.5 & -300.1 $\pm$ 0.9 \\
        D1 & 585.1 $\pm$ 1.3 & 579.9 $\pm$ 1.3 \\
        D2 & 640.6 $\pm$ 4.1 & 552.2 $\pm$ 2.1 \\ \hline
        AQ & 0 $\pm$ 3 & 0 $\pm$ 3 \\
        BQ & 87 $\pm$ 3 & -163 $\pm$ 3 \\
        CQ & 774 $\pm$ 3 & -257 $\pm$ 3 \\
        DQ & 610 $\pm$ 3 & 579 $\pm$ 3 \\
        Gal. Center & 451 $\pm$ 12 & 143 $\pm$ 5 \\ \hline
    \end{tabular}
    \caption{Image positions of the three quads (two radio and one optical) and the galaxy center of HS 0810+2554. The optical quad is an unresovled quadruply lensed RQQ (bottom; labelled AQ-DQ) whose image positions, along with the optical galaxy center, are taken from the CASTLES survey \citep{1998kochanek}. The two radio quads are two quadruply lensed radio jets (top; labelled A-D for pairs 1 and 2), where both jets originate from the RQQ seen in AQ-DQ. The image positions are centered on the brightest image (i.e., A) for each optical and radio positions. The lensing galaxy is at an estimated redshift of z$_l$ = 0.89 \citep{2011mosquera} and the three sources lie in the same source plane at z$_s$ = 1.51 \citep{2014chartas}.}
    \label{tab:positions}
\end{table}

In the classification of and endeavor to unify AGN morphologies, an important distinction to make about RQQs is that they are not radio silent. Bolstered by predictions from \citealt{2007white}, five gravitational lens systems known to have no radio detections down to $\sim$ 1mJy were found to have radio detections of a few tens of $\mu$Jy, including HS 0810+2554 \citep{2011jackson,2015jackson}. After discovering possible structure in the RQQ source, \citealt{2019hartley} re-observed HS 0810+2554 with e-MERLIN and EVN to discover the source has two nJy radio jets, which at the time were ``the faintest radio source[s] ever imaged." These sources manifest as eight images in the lens plane (blue and red in Figure \ref{fig:hs0810}). 

\begin{figure*}
    \centering
    \includegraphics[width=0.95\linewidth]{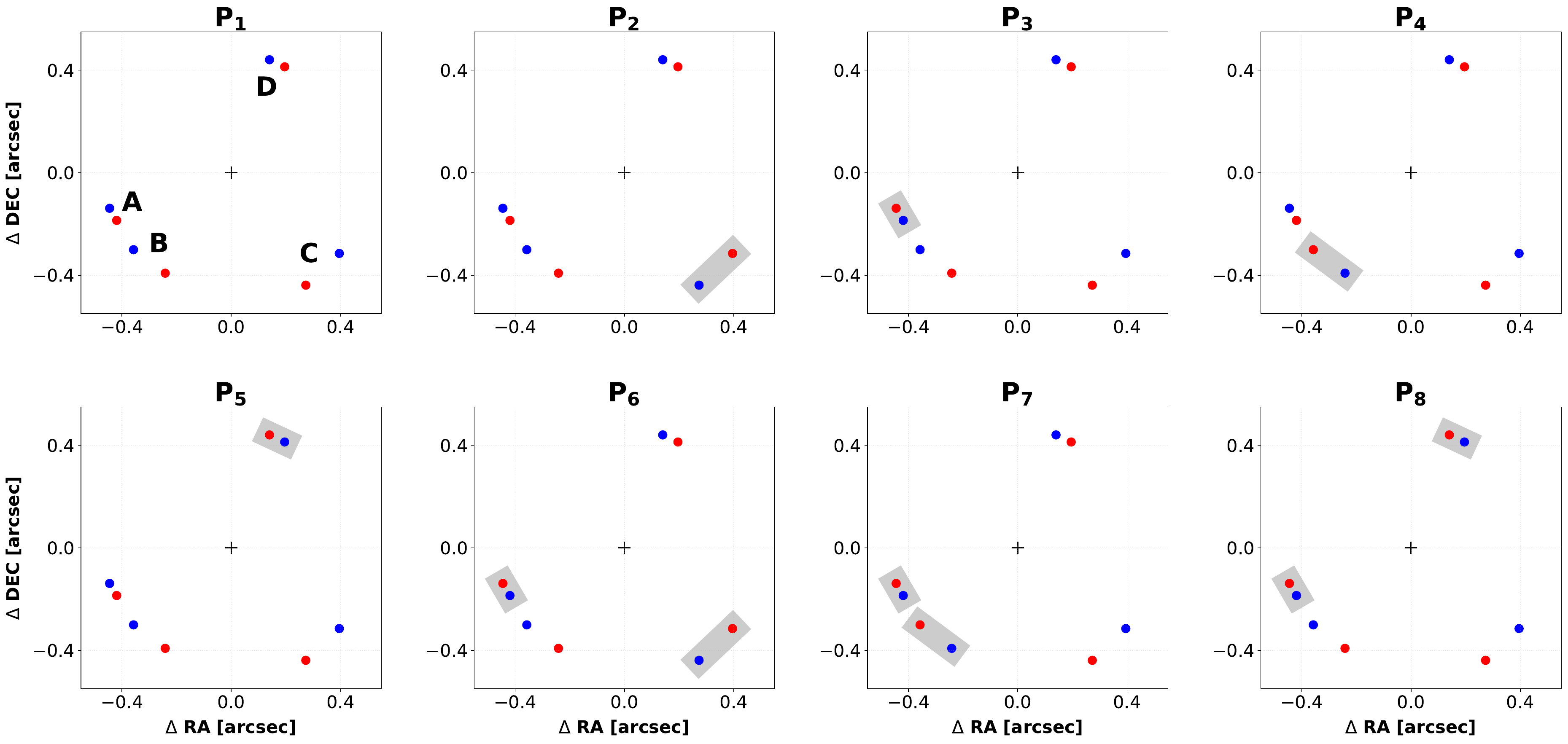}
    \caption{A pictorial representation of the eight unique 2x4 configurations (or pairings; P$_{1-8}$) of the eight radio images of HS 0810+2554 as outlined in Table \ref{tab:combs}. The four images in a configuration belonging to the first source ($\theta_{1,1-4}$) are colored blue and the second source ($\theta_{2, 1-4}$) colored red. Similarly to how images are underlined in Table \ref{tab:combs}, gray boxes are drawn around images to indicate when they are switched between sources. The image positions are reported in Table \ref{tab:positions} and centered on the CASTLES derived galaxy center with a (+6 mas, +4 mas) offset. A-D designates the relative brightness labelling scheme of the images.}
    \label{fig:combs}
\end{figure*}

The HS 0810+2554 image positions of all 12 images (four optical and eight radio) and the galaxy center are tabulated in Table \ref{tab:positions}. The image positions of the optical RQQ (labelled AQ-DQ) and galaxy center are taken from the CASTLES survey \citep{1998kochanek}. The eight radio image positions (labelled A-D for pairs 1 and 2) are taken from Table 5 of \citealt{2019hartley}\footnote{The HST image positions reported in Table 5 of \citealt{2019hartley} apply an offset of (-6 mas, -4 mas) to the image positions reported by the CASTLES survey \citep{1998kochanek}. However, the $\Delta$RA of images B \textit{HST} and C \textit{HST} are not consistent with this offset and should have been reported as 81.0 mas and 768.0 mas, respectively.}. The lensing galaxy was not observed in radio. Typically, lensing images are `paired' with their counter-images through spectroscopic identification, being assigned to the same background source. However, this could not be done with the eight radio images. The most plausible configuration of images was found by modelling the lensing galaxy with an SIE, iterating through all possible image configurations, and finding the best fit configuration. We denote this pairing of images as P$_1$ and is the configuration reported in Table \ref{tab:positions}. Because the images are not source identified, however, other image configurations are plausible based on lensing theory. In total, there exists eight possible unique 2x4 configurations (or pairings; P$_{1-8}$) of the eight radio images for one time ordering. These configurations are shown in the top half of Table \ref{tab:combs} and in Figure \ref{fig:combs}.

The 12 images are labelled based on their relative brightness (i.e., A-D) and, because no time-delay information exists, can be re-labelled based on their morphological time ordering \citep{2003saha}. All three quads are fold configurations with sources nearby each other in the source plane, and images nearby each other in the lens plane (i.e., all three A images, etc.), making it likely they all follow the same time ordering. The optical image C (CQ) is furthest from the galaxy's light center. Thus, the most probable time ordering for the three quads is C as first arriving, A second, B third, and D fourth, and is arrival sequence for image configurations P$_{1-8}$. Due to the distribution of images being near-circular, we additionally investigate the possibility of the reverse time ordering of the radio quads (i.e., D are first arriving, etc.) and designate the image configurations with primes P$_{1-8}^\prime$. The reverse time ordering is unlikely, but is included for a comprehensive and exhaustive analysis. These are the only two realistic time orderings. In total, the ambiguity of the eight radio images and two potential time orderings creates 16 unique states that could plausibly represent the true nature of the lensing system.

\begin{table}
    \centering
    \begin{tabular}{ccccccccc} \hline
        Pair & $\theta_{1,1}$ & $\theta_{1,2}$ & $\theta_{1,3}$ & $\theta_{1,4}$ & $\theta_{2,1}$ & $\theta_{2,2}$ & $\theta_{2,3}$ & $\theta_{2,4}$ \\ \hline
        P$_1$ & C1 & A1 & B1 & D1 & C2 & A2 & B2 & D2 \\
        P$_2$ & \underline{C2} & A1 & B1 & D1 & \underline{C1} & A2 & B2 & D2 \\
        P$_3$ & C1 & \underline{A2} & B1 & D1 & C2 & \underline{A1} & B2 & D2 \\
        P$_4$ & C1 & A1 & \underline{B2} & D1 & C2 & A2 & \underline{B1} & D2 \\
        P$_5$ & C1 & A1 & B1 & \underline{D2} & C2 & A2 & B2 & \underline{D1} \\
        P$_6$ & \underline{C2} & \underline{A2} & B1 & D1 & \underline{C1} & \underline{A1} & B2 & D2 \\
        P$_7$ & C1 & \underline{A2} & \underline{B2} & D1 & C2 & \underline{A1} & \underline{B1} & D2 \\
        P$_8$ & C1 & \underline{A2} & B1 & \underline{D2} & C2 & \underline{A1} & B2 & \underline{D1} \\ \hline
        P$_1^\prime$ & D1 & B1 & A1 & C1 & D2 & B2 & A2 & C2 \\
        P$_2^\prime$ & D1 & B1 & A1 & \underline{C2} & D2 & B2 & A2 & \underline{C1} \\
        P$_3^\prime$ & D1 & B1 & \underline{A2} & C1 & D2 & B2 & \underline{A1} & C2 \\
        P$_4^\prime$ & D1 & \underline{B2} & A1 & C1 & D2 & \underline{B1} & A2 & C2 \\
        P$_5^\prime$ & \underline{D2} & B1 & A1 & C1 & \underline{D1} & B2 & A2 & C2 \\
        P$_6^\prime$ & D1 & B1 & \underline{A2} & \underline{C2} & D2 & B2 & \underline{A1} & \underline{C1} \\
        P$_7^\prime$ & D1 & \underline{B2} & \underline{A2} & C1 & D2 & \underline{B1} & \underline{A1} & C2 \\
        P$_8^\prime$ & \underline{D2} & B1 & \underline{A2} & C1 & \underline{D1} & B2 & \underline{A1} & C2 \\ \hline
    \end{tabular}
    \caption{The 16 unique 2x4 configurations (or pairings) of the eight radio images of HS 0810+2554 for two time orderings: morphological (top; P$_{1-8}$) and its reverse (bottom; P$_{1-8}^\prime$). A pictorial representation of these configurations can be seen in Figure \ref{fig:combs}. When discovered, \citealt{2019hartley} found the configuration P$_1$ to be favored through lens modelling, not through source identification. The first subscript of $\theta_{\#,\#}$ denotes which source the grouping of four images belongs to and the second subscript denotes the time ordering (i.e., the first arriving image for the first source is denoted $\theta_{1,1}$). For clarity, images are \underline{underlined} when switched between source one and two.}
    \label{tab:combs}
\end{table}

Lens modelling with all 12 images as model constraints requires either the optical or radio image positions to be translated into the other's coordinate reference frame. However, there exists a systematic uncertainty between the two reference frames because the lens is not detected in the radio, leading to a few 10s of milliarcsecond (mas) uncertainty. As a result, we choose to omit the optical images from lens modelling and only consider the eight radio images, consistent with \citealt{2019hartley} and \citealt{2023amruth}. The eight images are centered on the CASTLES galaxy center with a nominal offset of (+6 mas, +4 mas). In other words, the center of the radio images is set to be (445 mas, 139 mas) with respect to image A1. Following the methodology of \citealt{2023amruth}, the measured radio flux ratios are also excluded due to the likelihood that the images fluxes were not fully recovered. Therefore, our investigation will determine if elliptical CDM mass models with higher order perturbations can reliably fit the eight image positions of the lensed radio jets. The close proximity of the radio images in the lens plane ($\approx$ 62 mas for images D1 and D2) in conjunction with the high precision astrometry ($\approx$ 1 mas) provides a unique opportunity to test CDM at these scales.

\section{Mass Models} \label{sec:mods}

To lowest order, the mass distribution of a lensing galaxy can be approximated as an elliptical power-law (EPL), where the surface mass density is given by $\kappa(r) \propto r^{-\alpha}$ \citep{1993kassiola,1994kormann}. The combined dark matter and baryonic mass profiles are generally well described by an EPL with a radial slope of $\alpha=1$, denoted the SIE \citep[][]{1994kormann,2003rusin,2004treu}. By themselves, however, these simply parameterized smooth lens models often do not reliably reconstruct lensing observables, and in these instances rely on the addition of `external' shear \citep{1997keeton,1997witt}. External shear can account for perturbations from nearby massive objects (i.e., galaxies, galaxy groups, etc.) that are necessary to include in modelling to reproduce lensing observables. However, external shear can also be partially accounting for lack of internal complexity in the mass model, like boxiness, diskiness, triaxiality, and lopsidedness \citep{1997keeton,2024etherington}. Even with the help from external shear, the SIE+EX or EPL+EX mass models often do not accurately recover lensing observables \citep{2004biggs,2006claeskens,2008jackson,2012sluse,2019shajib,2020wagner}. In these instances, the lensing galaxy is likely not dynamically relaxed and azimuthal perturbations beyond ellipticity, like boxiness, might be needed.

In the case of HS 0810+2554, the images of the RQQ in optical can and have been well fit by simply parameterized models. On the other hand, the eight radio images cannot be fit within their astrometric uncertainties with these models. In Section \ref{sec:existing} we briefly discuss the previous HS 0810+2554 mass model fits and their parameters, if given by the original paper. In Section \ref{sec:ours} we introduce the four mass models that we use to fit the eight radio image positions.
 
\subsection{Existing Models} \label{sec:existing}

The majority of existing HS 0810+2554 lens models were completed before the discovery of the eight radio images and thus only reconstruct the singular optical quad. Across these models the lensing galaxy has been continuously found to be well described with the SIE+EX model. Below is a brief description of these previous models. Unless specified, the best-fit mass models are SIE+EX. 

The first published model is from \citealt{2011assef} that used \texttt{lensmodel} \citep{2001keeton} to derive image magnifications. Although there was no direct mention of the model's fitness or a discussion of the best-fit model parameters, it is presumed to have been well fit. Later, observations of the system in X-ray revealed a nearby galaxy group in projection \citep{2014chartas,2016chartas}. From 5 ks and 100 ks observations the external shear contribution from this group was estimated to be $\gamma = 0.016 \pm 0.003$ and $\gamma = 0.026 \pm 0.003$, respectively. With the external shear constrained by these measurements, \citealt{2014chartas,2016chartas} modelled the X-ray image positions of the main RQQ with \texttt{glafic} \citep{2010oguri}. \citealt{2020chartas} repeated this methodology to investigate the mm-continuum of the source galaxy, but instead used the optical image positions from CASTLES \citep{1998kochanek}. The predicted magnifications from all three models were consistent with \citealt{2011assef}. Initial radio observations of HS 0810+2554 found four images of unresolved radio emission \citep{2015jackson}. The best-fit model was nearly circular with a moderate contribution from external shear $\gamma = 0.023 \pm 0.006$, consistent with the shear prediction from the nearby galaxy group. After being re-observed in optical, \citealt{2020nierenberg} found a best fit elliptical power-law model with moderate ellipticity and nearly twice the contribution from external shear than predicted. 

All of the previously discussed smooth mass models were able to successfully model the four optical images of HS 0810+2554 with no difficulty. However, the discovery and subsequent modelling of the eight radio images by \citealt{2019hartley} clearly indicate that the lens galaxy is more complicated than previously thought. Their best-fitting SIE+EX model was not able to successfully reproduce the eight image positions within astrometric uncertainty ($\chi=7.05$) even though the model parameters were in reasonable agreement with \cite{2015jackson}. An acceptable fit of $\chi^2_\textup{red}=1$ was only achieved by relaxing the $\approx 1$ mas astrometric uncertainties to be eight times greater for nearly all images. 

To test the small-scale effects of FDM, \citealt{2023amruth} added FDM perturbations with a de Broglie wavelength of $\lambda_{dB}$ = 180 pc in the form of Gaussian random fields (GRFs) to a best-fitting elliptical power-law model. Although not direct evidence for FDM, \citealt{2023amruth} found that FDM is able to create perturbations on scales that encompass the average positional anomaly between the observed image positions and the predicted image positions from their best-fit elliptical power-law mass model. That is to say, FDM can create perturbations on the scale necessary to fix the positional anomalies. The fact that smooth CDM mass models cannot sufficiently fit the eight radio images within astrometric uncertainties, and the ability of small-scale perturbations to correct the position anomalies of smooth mass models, indicate that more complex CDM mass models (i.e., lopsidedness, boxiness, etc.) need to be investigated.

\subsection{Our Models} \label{sec:ours}

The primary addition of our mass modelling is the inclusion of azimuthal deviations from ellipticity in the form of Fourier perturbations, or multipoles. Elliptical galaxies, especially those at higher redshifts (i.e., HS 0810+2554), are often perturbed by mergers or asymmetric mass accretion. The effects of these interactions are most commonly known to be encoded in the multipoles displayed in a galaxy's stellar isophotes \citep{2005khochfar,2006hao,2014chaware,2017mitsuda}. The most commonly discussed of these multipoles are the third- and fourth-order deviations ($m=3,4$), where the fourth order perturbations are labeled `boxiness' ($\theta_4=45\degree$) and `diskiness' ($\theta_4=0\degree$) for non-negative $a_4$. For completeness, we label the third-order $m=3$ as `triangleness'. The absence of these perturbations in lens modelling have been shown to lead to a biased view of lensing galaxies \citep{2022vyvere1,2024cohen,2024oriordan}. 

A less talked about multipole, the $m=1$ multipole, or lopsidedness, is the result of tidal interactions from recent or ongoing mergers and nearby galactic companions non-uniformly shifting the centers of isodensity contours \citep{2024amvrosiadis}. The inclusion of lopsidedness has been necessary to precisely fit certain observed lenses \citep{2020zegeye,2021barrera} and help explain the deviation from ellipticity displayed in current population of quads \citep{2018gomer,2024miller}. These three multipoles $m=1,3,4$ are included in our analysis to investigate if their physically motivated small-scale perturbations are enough to explain the eight radio images, or if other avenues are needed.

Our analysis consists of fitting the eight radio image positions to four mass models of increasing complexity (Mod$_{1-4}$) that are combinations of three different mass model components: the elliptical power-law (Section \ref{sec:epl}), multipoles (Section \ref{sec:multipoles}), and external shear (Section \ref{sec:ex}). To create a baseline, our first model Mod$_{1}$ consists only of an elliptical power-law and a constant external shear of magnitude $\gamma$ = 0.026 and $\theta_{\gamma} = 91.5 \degree$, consistent with observations from \citep{2016chartas}. This constant external shear is applied to all four models. The second model Mod$_2$ includes lopsidedness ($m=1$) and isolates the contributions from dark matter and baryonic matter by modelling them as two separate EPL profiles. The third model Mod$_3$ includes lopsidedness, triangleness, boxiness, and diskiness with the multipoles $m=1,3,4$, and models the dark matter and baryonic matter separately. Finally, our last model Mod$_4$ is the same as Mod$_3$ but includes an external shear component that is allowed to vary. Table \ref{tab:models} itemizes each of the four models, including the parameters of each mass profile and the total number of free parameters. The degrees of freedom for Mod$_3$ and Mod$_4$ are negative. As will be presented in Section \ref{sec:results}, simpler models (i.e., Mod$_1$ and Mod$_2$) do not reconstruct the image position well, motivating the use of more complex models.

\begin{table}
\setlength{\tabcolsep}{5.5pt}
    \centering
    \begin{tabular}{c|cccccccc} \hline
        Profile & \multicolumn{4}{c}{Parameters} & Mod$_1$ & Mod$_2$ & Mod$_3$ & Mod$_4$ \\ \hline
        EPL$_{1}$ & $t_1$ & $b_1$ & $q_1$ & $\varphi_1$ & \checkmark & \checkmark & \checkmark & \checkmark \\
        EPL$_{2}$ & $t_2$ & $b_2$ & $q_2$ & $\varphi_2$ & --- & \checkmark & \checkmark & \checkmark \\
        M1 & $a_1$ & $\theta_1$ & ... & ... & --- & \checkmark & \checkmark & \checkmark \\
        M3 & $a_3$ & $\theta_3$ & ... & ... & --- & --- & \checkmark & \checkmark \\
        M4 & $a_4$ & $\theta_4$ & ... & ... & --- & --- & \checkmark & \checkmark \\
        EX & $\gamma$ & $\theta_{\gamma}$ & ... & ... & --- & --- & --- & \checkmark \\ \hline
        EX & \multicolumn{2}{c}{$\gamma=0.026$} & \multicolumn{2}{c}{$\theta_{\gamma}=91.5\degree$}& \checkmark & \checkmark & \checkmark & \checkmark \\ \hline
        \multicolumn{5}{c}{Total \# of Free Parameters (+6)$^\textbf{**}$} & 10 & 16 & 20 & 22 \\ \hline
    \end{tabular}
    \caption{Summary of our four main mass models (Mod$_{1-4}$) and parameters. The parameters for the elliptical power-law (EPL), multipoles (M\#), and external shear (EX) profiles are described in Sections \ref{sec:epl}, \ref{sec:multipoles}, and \ref{sec:ex}, respectively. All models have a non-varying external shear consistent with the nearby galaxy group \citep{2016chartas}. $\textbf{**}$All models have an additional six free parameters: two varying terms for the galaxy center (x$_g$, y$_g$) and four for the two source positions (x$_{s1}$, y$_{s1}$, x$_{s2}$, y$_{s2}$). The centers of EPL$_{1}$ and EPL$_{2}$ are co-aligned.}
    \label{tab:models}
\end{table}

The eight radio images are centered on the HST galaxy center and with a nominal offset of (+6 mas, +4 mas) to ensure A1 is not coincident with AQ. The center of the lensing galaxy (x$_g$, y$_g$) is initialized at (0,0) and allowed to vary. For models Mod$_{2-4}$, the two EPL profiles are co-aligned and walk around synchronously. The two ellipticities and position angles are not tied to each other and are allowed to vary freely. The baryonic matter normalization and slope parameters, $b_2$ and $t_2$, are restricted to be greater than that of the dark matter to have a higher central normalization $b_2 \geq b_1$ and to fall off quicker $t_2 \geq t_1$.  Furthermore, the slopes of all EPL profiles are restricted between $0.6 \leq t \leq 1.4$ and ellipticities restricted between $0.5 \leq q \leq 1.0$. Multipole normalizations $a_1$, $a_3$, and $a_4$ are restricted to be positive.

The modelling software was written in \texttt{Python} and utilizes the `Nelder–Mead' minimization method from the package \texttt{Scipy} to fit the eight radio image positions. The figure-of-merit for a given model is given by a modified $\chi$ value defined as

\begin{equation} \label{eq:chi}
    \chi = \sqrt{\chi^2} = \sqrt{\frac{1}{2N_{\rm{ims}}} \sum_{i=1}^{2} \sum_{j=1}^{4} \left( \frac{(x_{ij}-x^{\rm{\prime}}_{ij})^2}{\sigma_{x,ij}^2} + \frac{(y_{ij}-y^{\rm{\prime}}_{ij})^2}{\sigma_{y,ij}^2} \right) }
\end{equation}

\noindent where $i$ delineates between the two sources, $j$ the arrival time order (i.e., $j$=1 is first arriving), $x$ and $y$ are the image positions of the observed data in the lens plane, $x^\prime$ and $y^\prime$ the model image positions, $\sigma$ the astrometric errors, and $N_{\rm{ims}}$ the observed number of images ($N_{\rm{ims}}=8)$. Randomized initial simplexes are generated from parameter ranges that span a breadth of realistic values. For multipole normalizations ($a_1$, $a_3$, and $a_4$), the initial simplex is given the range (0, 0.04) to explore and can walk to larger values outside this range if it desires.

After finding a solution the program iterates upon the solution multiple times to help overcome potential local minima. Image configurations are enforced by calculating the arrival sequence of the model images and fitting each model image to its corresponding image in the observed arrival sequence. For example, in image configuration P$_1$ the first arriving model image will always be compared to C1 (or C2), regardless of its location. Each of the 16 image configurations for the eight radio images (Table \ref{tab:combs}) are fit by each of the four mass models, resulting in a total of 64 mass models. To our knowledge, no other strong lensing study has modelled a single lens as extensively.

\subsubsection{Elliptical Power-Law} \label{sec:epl}
As discussed before, the projected mass density profile of the dark matter and baryonic matter in an elliptical galaxy can be generally well modelled by an SIE or, in the more general case, an EPL profile. Elliptical lensing potentials, where the ellipticity of the lens is introduced in the potential itself, have the unfortunate consequence of becoming `peanut shaped' in their mass distribution at large ellipticities \citep{1993kassiola}. To counteract this effect we choose the popular EPL from \citealt{2015tessore} that incorporates the ellipticity at the level of the convergence, $\kappa$, which is given by 
\begin{equation} \label{eq:epl}
    \kappa(R) = \frac{2-t}{2} \left( \frac{b}{R} \right)^t,
\end{equation}

\noindent where 0 < $t$ < 2 is the slope of profile ($t=1$ for SIE), $b$ is the scale length (or the Einstein radius for a circular lens), and $R$ is the elliptical radius from the lens center defined by
\begin{equation} \label{eq:r}
    R = \sqrt{q^2x^2+y^2},
\end{equation}

\noindent where $q$ is the elliptical axis ratio. The position angle of the semi-major axis $\varphi$ is given by
\begin{equation} \label{eq:pa}
    \varphi = \arctan(qx,y) .
\end{equation}

One benefit of using this specific profile is that the lensing potential and deflection angles can be computed analytically (see Section 4 of \citealt{2015tessore}), saving computational time. Aside from the source positions, to model a lens with the EPL mass profile a total of six model parameters are needed: two for the galaxy center (x$_g$, y$_g$) and four for the EPL profile ($b$, $q$, $t$, and $\varphi$).

\subsubsection{Multipoles} \label{sec:multipoles}

There are multiple multipole definitions in strong lensing literature \citep{2022vyvere,2022fores,2024cohen,2024oriordan,2024gilman}. We chose to use the convention from \citealt{2016kawamata}. The lensing potential perturbation created by a multipole at $(r, \theta)$ in the lens plane is given by
\begin{equation} \label{eq:mult}
    \psi = -\frac{a_m}{m} r^2 \cos[m(\theta-\theta_m)]
\end{equation}

\noindent where $m$ designates the multipole number (i.e., $m=1,3,4$), $a_m$ the normalization or amplitude, and $\theta_m$ the position angle of the given multipole, which lies in the positive x-direction and moves counter-clockwise. The multipoles are aligned with the galaxy's center (i.e., the EPL center) and modelled with a total of two parameters, $a_m$ and $\theta_m$, per multipole $m$.

Our choice of a multipole lensing potential with $\psi \propto r^2$ is in contrast to other galaxy-scale analyses with $\psi \propto r$. We choose this formalism because galaxies are often observed to have multipole perturbations extending into their outer regions with increasing or constant magnitudes \citep{2014chaware}. Due to the inner regions of galaxies having shorter relaxation time scales, following a merger or a perturbation, one would expect the magnitude of perturbations to increase with radius, not decrease.

\subsubsection{External Shear} \label{sec:ex}

As shown by \citealt{2014chartas,2016chartas}, there is a non-zero shear contribution from a nearby local galaxy group. The lensing potential associated with external shear at $(r, \theta)$ in the lens plane is given by
\begin{equation} \label{eq:ex}
    \psi = \frac{\gamma}{2} r^2 \cos[2(\theta-\theta_{\gamma})],
\end{equation}

\noindent where $\gamma$ is the magnitude of the shear component and $\theta_{\gamma}$ the angular direction, which has a period of $\pi$ and lies along the positive and negative y-axis. The direction of the local group is estimated from Figure 12 of \citealt{2016chartas} to be $1.5\degree$ northwest of HS 0810+2554. Therefore, a constant external shear of $\gamma = 0.026$ at $\theta_{\gamma} = 91.5\degree$ is included in all models Mod$_{1-4}$. The last model Mod$_{4}$ includes an additional external shear term that is allowed to vary both $\gamma$ and $\theta_{\gamma}$.

\section{Results} \label{sec:results}

\begin{table}
    \centering
    \begin{tabular}{c|cccc} \hline
        & Mod$_{1}$ & Mod$_{2}$ & Mod$_{3}$ & Mod$_{4}$ \\ \hline
        \multirow{2}{*}{Pair} & \multirow{2}{*}{EPL}& EPL+BM & EPL+BM & EPL+BM \\
         & & +M1 & +M134 & +M134+EX \\ \hline
        P$_1$ & \textbf{8.58}& 5.33 & 4.20 & 3.35 \\
        P$_2$ & 32.67 & 20.65 & 18.04 & 18.50 \\
        P$_3$ & 12.01 & \textit{\textbf{4.70}} & \textit{\textbf{3.79}} & \textbf{2.95} \\
        P$_4$ & 24.45 & 9.83 & 7.29 & 7.12 \\
        P$_5$ & 11.03 & 9.31 & 6.62 & 5.47 \\
        P$_6$ & 24.62 & 9.56 & 5.39 & 5.81 \\
        P$_7$ & 32.93 & 20.65 & 18.38 & 18.59 \\
        P$_8$ & 14.13 & 8.74 & 6.39 & 5.53 \\ \hline
        P$_1^\prime$ & \textit{\textbf{8.18}} & \textbf{5.20} & \textbf{4.59} & \textit{\textbf{1.59}} \\
        P$_2^\prime$ & 31.38 & 18.98 & 19.80 & 19.23 \\
        P$_3^\prime$ & 8.94 & 5.67 & 4.77 & 2.03 \\
        P$_4^\prime$ & 23.84 & 10.79 & 10.82 & 8.98 \\
        P$_5^\prime$ & 12.46 & 9.44 & 7.55 & 5.83 \\
        P$_6^\prime$ & 24.28 & 9.81 & 9.98 & 7.90 \\
        P$_7^\prime$ & 31.35 & 19.46 & 19.00 & 18.57 \\
        P$_8^\prime$ & 12.22 & 9.50 & 7.17 & 5.43 \\ \hline
    \end{tabular}
    \caption{The lowest chi values (see Eq. \ref{eq:chi}) for each of the eight image configurations (or pairings; P$_{1-8}$) and two time orderings (see Table \ref{tab:combs}). For each time ordering (top and bottom), the best-fit configuration for each model is in \textbf{bold}. The best overall model per column is \textit{italicized}. The abbreviation BM stands for Baryonic Matter, which is modelled with an EPL profile.}
    \label{tab:res}
\end{table}

The best-fit $\chi$ values for each of the 16 image configurations and four mass models are listed in Table \ref{tab:res}. For un-primed image configurations (morphological arrival sequence), P$_1$ was favored for model Mod$_1$ (EPL + constant EX), reproducing the results from \citealt{2019hartley}. However, the reverse time ordering P$_1^\prime$ was found to have a slightly better $\chi$ value than P$_1$ for Mod$_1$. For models Mod$_{2-4}$ and un-primed configurations, P$_3$ had the lowest $\chi$ values with P$_1$ being a close second. For primed image configurations (reversed arrival sequence), P$_1^\prime$ was found to have the lowest $\chi$ values for each of the four models, with P$_3^\prime$ being a close second. The overall best-fitting model with the lowest $\chi$ value of all 64 models was Mod$_4$--P$_1^\prime$ with $\chi=1.59$, providing a good fit to the eight radio image positions and a physical looking $\kappa$ map. 

\begin{figure*}
    \centering  
    \includegraphics[width=\textwidth]{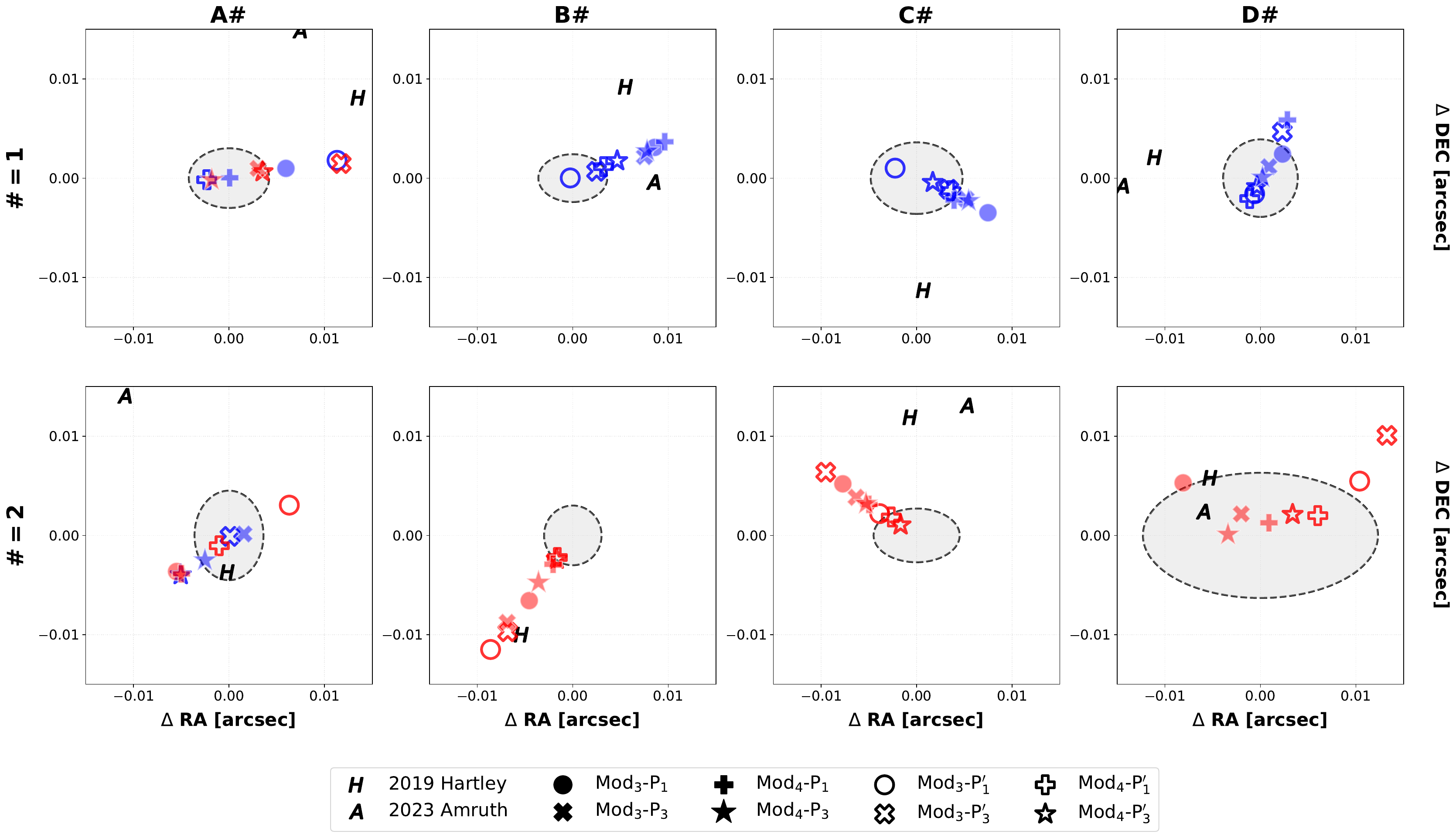}
    \caption{A zoomed in view centered around each of the eight observed radio image positions, with 3$\sigma$ error ellipses and the predicted positions from eight of our best fitting models. The top figures are centered on the image positions of the `first' P$_1$ quad (\#=1) and the bottom is centered on the `second' P$_1$ quad (\#=2; see Table \ref{tab:positions} and P$_1$ in Table \ref{tab:combs}). The columns are ordered alphabetically (A-D), not based on arrival time. The eight best-fit models are Mod$_{3,4}$ for image configurations P$_{1,3}$ and P$_{1,3}^\prime$. The marker shape for each image configuration is the same for both time orderings, with the reverse time order being unfilled, i.e., Mod$_3$--P$_1$ (filled circle) and Mod$_3$--P$_1^\prime$ (unfilled circle). Model images belonging to the `first' quad within its image configuration are colored blue and `second' quad colored red. The images of A1 and A2 are red and blue, respectively, for P$_3$ because those images are switched when compared to P$_1$ (see Table \ref{tab:combs} and Figure \ref{fig:combs}). The image positions from the best fitting SIE+EX from \citealt{2019hartley} ($\chi=7.05$) are included as H's for comparison and the best fitting EPL model from \citealt{2023amruth} ($\chi=9.84$) as A's. Images C1 and B2 from \citealt{2023amruth} fall outside the window. }
    \label{fig:best}
\end{figure*}

\begin{figure*}
    \centering
    \includegraphics[width=0.9\linewidth]{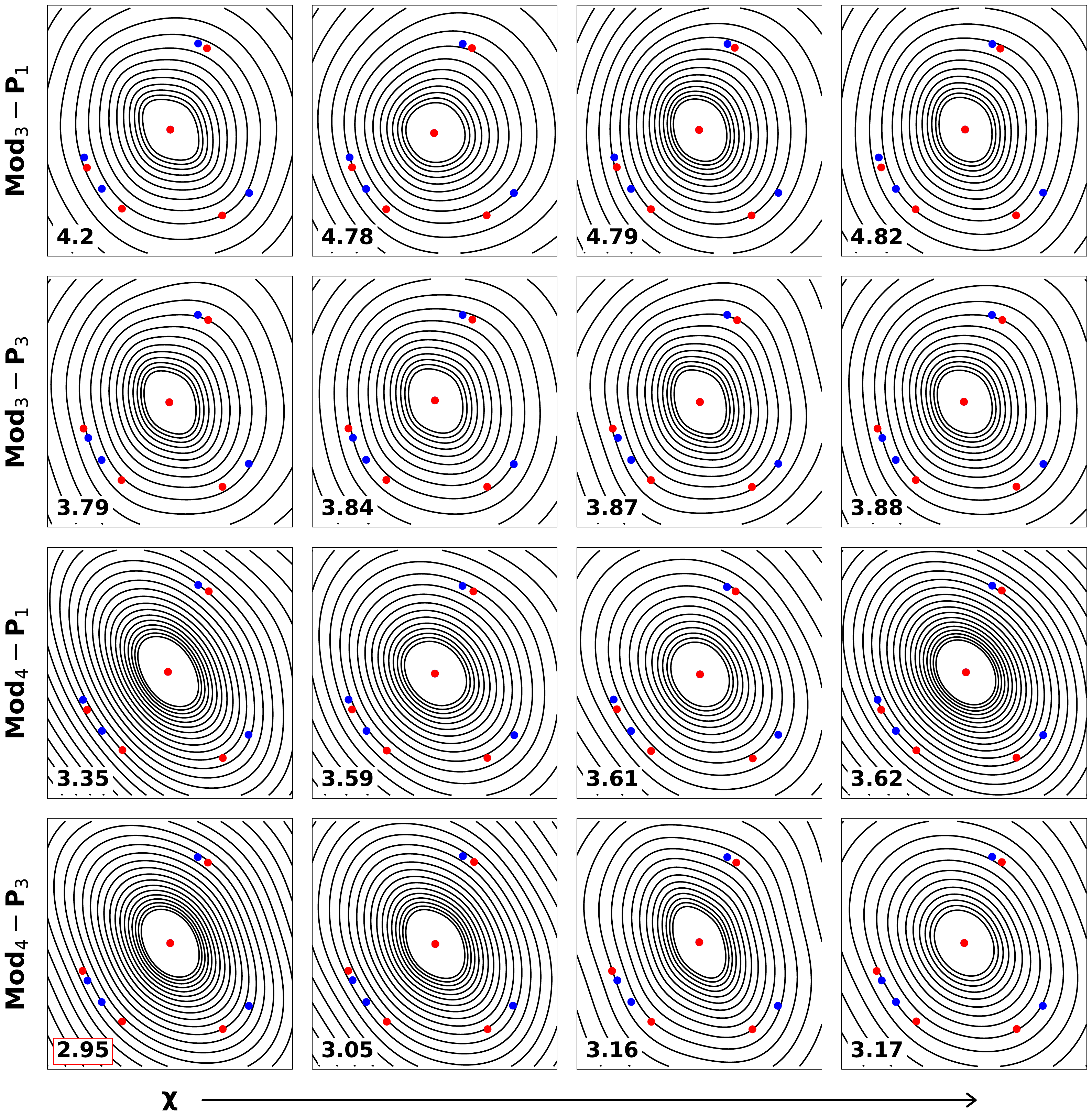}
    \caption{Convergence, or $\kappa$, maps for the four best-fitting solutions of the two models Mod$_{3-4}$ and image configurations P$_1$ and P$_3$. Column 1 displays the $\kappa$ maps with the lowest $\chi$ values, whose positional errors can be seen in Figure \ref{fig:best}. Columns 2-4 show the three next best solutions. The fit parameters for all of these models are tabulated in Table \ref{tab:params}. The reverse time ordering $\kappa$ maps are displayed in Figure \ref{fig:kappas2}. The $\kappa$ map highlighted with the red rectangle ($\chi$=2.95) is the best-fit model with the morphological time ordering. Additionally, we consider this model to be reasonably physical, whereas other models displayed are less likely to be physical.}
    \label{fig:kappas1}
\end{figure*}

\begin{figure*}
    \centering
    \includegraphics[width=0.9\linewidth]{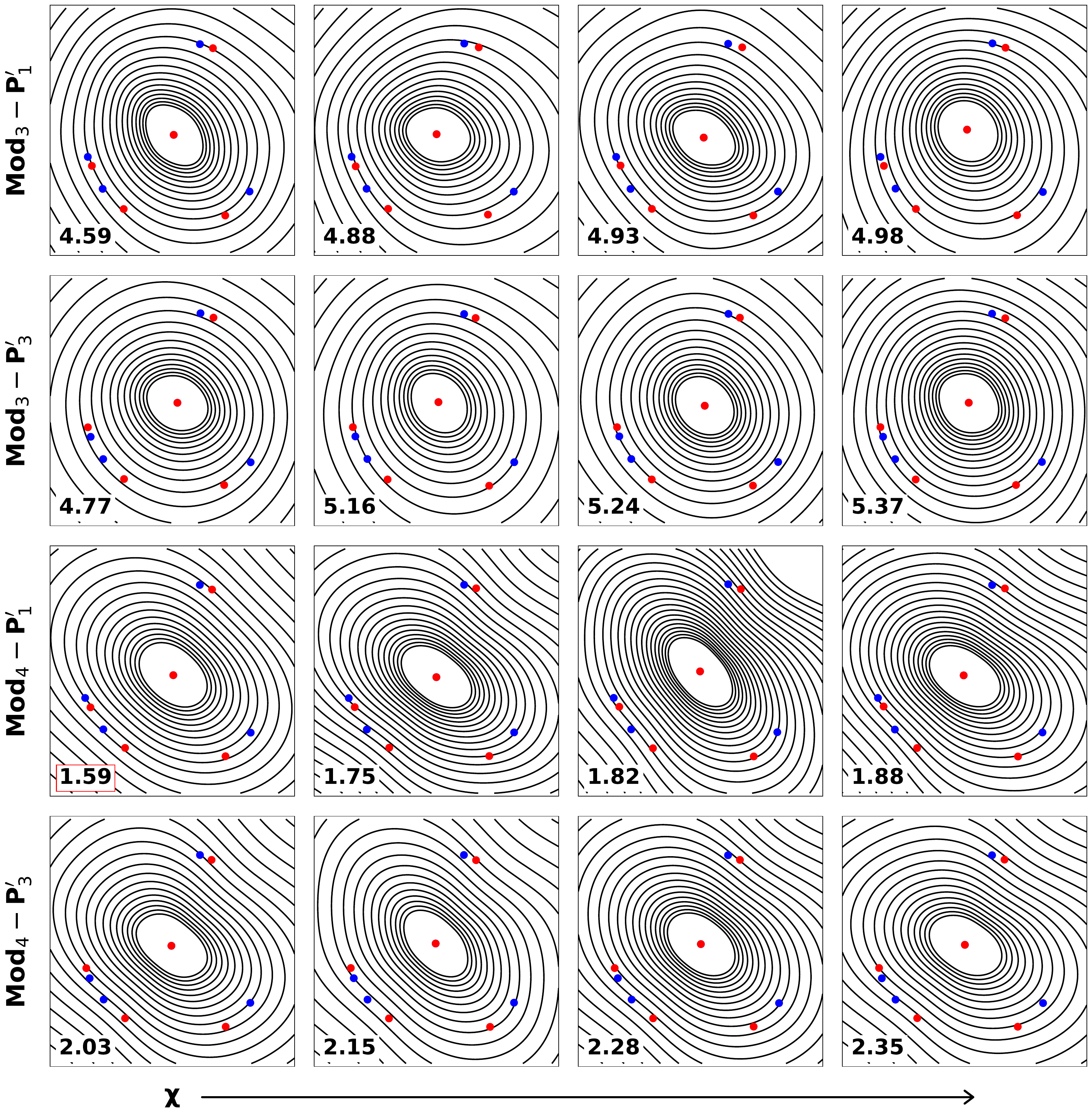}
    \caption{Convergence, or $\kappa$, maps for the four best-fitting solutions of the two models Mod$_{3-4}$ and image configurations P$_1^\prime$ and P$_3^\prime$. Column 1 displays the $\kappa$ maps with the lowest $\chi$ values, whose positional errors can be seen in Figure \ref{fig:best}. Columns 2-4 show the three next best solutions. The fit parameters for all of these models are tabulated in Table \ref{tab:params}. The standard time ordering $\kappa$ maps are displayed in Figure \ref{fig:kappas1}. The $\kappa$ map highlighted with the red rectangle ($\chi$=1.59) is the best-fit model with the reverse time ordering. Additionally, we consider this model to be reasonably physical, whereas other models displayed are less likely to be physical.}
    \label{fig:kappas2}
\end{figure*}

Figure \ref{fig:best} provides a zoomed-in view around the observed image positions showing the model predicted image positions from eight best-fit models: Mod$_{3,4}$ for image configurations P$_{1,3}$ and their reverse time orderings P$_{1,3}^\prime$. The parameters for these models are tabulated in Table \ref{tab:params}. The best-fit smooth mass models from \citealt{2019hartley} ($\chi=7.05$) and \citealt{2023amruth} ($\chi=9.84$) are shown alongside our best-fit mass models. The $\kappa$ maps for the best-fit Mod$_{3,4}$--P$_{1,3}$ models are shown in Figure \ref{fig:kappas1} (Column 1), alongside their three next best fitting solutions (Columns 2-4). The similar $\kappa$ maps for Mod$_{3,4}$--P$_{1,3}^\prime$ models are shown in Figure \ref{fig:kappas2}. We do not consider all models shown in Figures \ref{fig:kappas1} and \ref{fig:kappas2}, as not all models are necessarily realistic looking (i.e., display abnormal warping or extreme undulations; see Figure \ref{fig:kappas2} Mod$_4$--P$_1^\prime$). Models that are less likely to be physical are included to display the range of degenerate galaxy morphologies. The two models that we consider are reasonably physical are $\chi=1.59$ Mod$_4$--P$_1^\prime$ in Figure \ref{fig:kappas2} and $\chi=2.95$ Mod$_4$--P$_3$ in Figure \ref{fig:kappas1} (highlight by red rectangles). The slope of the 2D-$\kappa$ radial profile around the image radius for these two models is approximately isothermal.

The $\kappa$ map of the best-fit Mod$_4$--P$_1^\prime$ model has an elliptical shape near the center of the galaxy, which then becomes boxy at the outer radii near the radius of the image positions. The three multipoles have amplitudes of $a_1=8.5\times10^{-3} \pm ~3\times10^{-4}$, $a_3=4.0\times10^{-3} \pm ~7\times10^{-5}$, and $a_4=5.8\times10^{-3} \pm ~8\times10^{-5}$, on the order of typical values seen in literature \citep[$a_m \lesssim 10^{-2}$][]{}. For the best-fit  Mod$_4$--P$_3$ model, the multipoles have amplitudes of $a_1=5.1\times10^{-7} \pm ~2\times10^{-8}$, $a_3=5.3\times10^{-3} \pm ~1\times10^{-4}$, and $a_4=4.5\times10^{-3} \pm ~8\times10^{-5}$. The uncertainties are derived with an MCMC utilizing the \texttt{Python} package \texttt{emcee} \citep{2013Foreman-Mackey}. The strength of the multipole in $\kappa$ is constant, having no radial dependence, allowing for stronger perturbations at larger radii than other models. Due to the center of the galaxy being dominated by the two EPL components, the Fourier perturbations are less noticeable at smaller radii and become more pronounced at larger radii. This creates a radial-gradient in the multipole perturbations, similar to galaxies seen in \cite{2014chaware}.  

In general, all eight models shown in Figure \ref{fig:best} were able to fit images A1, A2, D1, and D2. The other images, B1, B2, C1, and C2 were systematically offset in a constant direction from the observed image location. With the $+$x-axis as the origin, the model predicted images for B1 and B2 are offset at $\theta_\textup{B1} \approx 20\degree$ and $\theta_\textup{B2} \approx 235\degree$, respectively. Model predicted images C1 and C2 are offset at $\theta_\textup{C1} \approx -30\degree$ and $\theta_\textup{C2} \approx 145\degree$, respectively. Switching the images in B and C between the two sources, which is equivalent to image configuration P$_8$, does not change this trend and results in the same azimuthal offsets. These four images having a clear preferential direction in their positional anomalies could indicate missing mass near the location of these images (i.e., subhalos) or potential issues with the astrometry, the latter of which we discuss in Section \ref{sec:discussion}. We note that initial, preliminary testing with an EPL profile and two subhalos were not able to successfully reproduce the image positions alone.

The angular separation of the two radio jets sources for the overall best-fitting model Mod$_4$--P$_1^\prime$ was found to be 6.6 mas. The separation corresponds to a physical separation of 57.6 pc in the source plane, with a fiducial cosmology of H$_0 = 70.0$, $\Omega_{m}=0.27$, $\Omega_\Lambda=0.73$. This source separation is consistent with \citealt{2023amruth} Extended Data Figure 5, which shows a source separation of approximately 6 mas, and only slightly smaller the 8.2 mas reported by \citealt{2019hartley} Table 4. Overall, the primed configurations preferred source separations slightly larger than the un-primed configurations. The latter of which found a source separation of 4.1 mas for Mod$_4$--P$_1$ and 4.6 mas for Mod$_4$--P$_3$. 

If the two radio sources originate from the quasar, then the quasar should lie between the two radio sources in the source plane. We investigate this for our two best models Mod$_4$--P$_1^\prime$ and Mod$_4$--P$_3$ in Appendix \ref{sec:qq}. With the lens parameters fixed, the best fitting quasar source position is determined by comparing its model images to the observed optical image morphology, not exact image positions. Figure \ref{fig:qq} shows the two best-fit models. Both models are physical as the quasar falls between the two radio source positions in the source plane.

In the lens plane, the center of the lensing galaxy was allowed to walk away from (0,0) during fitting. The model Mod$_4$--P$_1^\prime$ found a lens center that is quite far from the origin at (3.4 mas, -22.7 mas), or approximately 180 pc away. In general, the lens center for primed configurations significantly deviated from the origin ($\sim 20-30$ mas). Whereas the un-primed configurations typically stuck nearby the origin ($\sim 0-10$ mas).  Without multipoles, the best solutions for Mod$_1$--P$_1$ found the lens center to be far away at approximately (26 mas, 6 mas) and Mod$_1$--P$_1^\prime$ at (32 mas, -21 mas). The inclusion of higher order perturbations did allow for solutions with lens centers nearby the origin. For example, the Mod$_4$--P$_1^\prime$ model with $\chi=1.82$ as shown in Row 3, Column 3 of Figure \ref{fig:kappas2} only deviated from the origin by $\approx$ 4 mas.

\section{Discussion} \label{sec:discussion}

The inclusion of higher order perturbations in our CDM mass model fitting has decreased the image positional anomalies in the gravitational lens HS 0810+2554. Our best-fit model Mod$_4$--P$_1^\prime$ with $\chi = 1.59$ provides a good fit to the observed positions of the eight radio images. Interestingly, the image configuration is in the reverse arrival sequence (primed). The model Mod$_4$--P$_3$ was the best-fit morphological time ordering (un-primed) model with $\chi=2.95$. The model Mod$_4$ (and Mod$_3$) has negative degrees of freedom, which are reasonable to consider because simpler models (Mod$_1$ and Mod$_2$) failed, and galaxies can display deviations from ellipticity. Therefore, perturbations in the form of lopsidedness, triangleness, boxiness, and diskiness, i.e., $m=1,3,4$, do seem to be necessary in reproducing the image positions of the eight radio images. Many of our solutions display a radial-gradient in the strength of the three multipoles in $\kappa$, where the Fourier perturbations become more prominent at larger radii. The gradient itself does not stem from our choice of radial dependence in the multipole lensing potential, because when converted to $\kappa$ the multipole amplitude is constant as a function of radius (Eq. \ref{eq:mult}). Instead, the gradient stems from the suppression of multipoles at smaller radii by the two EPL profiles, which tapper off at larger radii, whereas the perturbations do not. 

Some of our models show both diskiness and boxiness in their $\kappa$ distributions. For example, the inner regions of the galaxy shown in Column 2, Row 4 of Figure \ref{fig:kappas2} ($\chi = 3.05$) are disky and become boxy at larger radii. Galaxies have been observed with similar properties \citep{2009kormendy}. \citealt{2014chaware} found isophotal shapes of galactic inner regions differ from those of their outer regions, suggesting they evolve somewhat independently and possibly as a consequence of a recent merger. Additionally, initial, preliminary tests (not presented here) indicated that an inverse radial dependence of multipoles in $\kappa$, where the strength of multipoles decrease with radius, were insufficient to model the radio images. This indicates that multipole perturbations likely persist in the outer regions of the galaxy. Thus, these findings could point to HS 0810+2554 having undergone a recent or on-going merger.

One of the motivations for the inclusion of reverse time ordering (lower half of Table \ref{tab:res}; primed) is that the galaxy center was not detected in the radio by \citealt{2019hartley} and there exists no measured time delays for HS 0810+2554 at any wavelength. Because the galaxy lens was not observed in radio, the arrival sequence of the radio images is presumed to follow the morphological arrival sequence of the optical quad. In most cases the morphological arrival sequence matches the sequence derived from time delays, as can be seen in \cite{2025dux}. However, the center of mass of the lensing galaxy need not be coincident with the center of light, particularly in galaxies affected by recent or on-going mergers. Due to the near circular nature of the image distribution in the plane of the sky, the galaxy center would only need to move a few tens of mas in the ($+\Delta$RA, $-\Delta$DEC) direction to reverse the time ordering. This positional offset is observed in our primed models. For example, the best-fit lens center (i.e., the location of the EPL) for Mod$_1$--P$_1^\prime$ models were located at approximately ($+6$ mas, $-27$ mas) with respect to the lens galaxy center found with Mod$_1$--P$_1$. By obtaining time delays---at any wavelength---the degeneracy between the two arrival sequences can be broken. Time delays of the eight radio images are likely not obtainable for HS 0810+2554 \citep{2014gurkan,2015rumbaugh}. Instead, optical observations of the lens by programs like COSmological MOnitoring of GRAvItational Lenses \citep[COSMOGRAIL;][]{2005Eigenbrod} or Time Delay COSMOgraphy \citep[TDCOSMO;][]{2020millon,2025dux} could be used to obtain time delays, which are likely to be similar to those in the radio. Not only would time delays break the degeneracy, but they would also provide more lens modelling constraints.

With that being said, the flux ratios of the four quasar images in IR provides additional evidence that the images likely follow the morphological arrival sequence. The relative brightness of images can be indicative of their arrival time, although not always: 2nd arriving (minima) is typically brighter than 3rd (saddle) \citep{2002schechter,2011saha}, and 1st (minima) is typically brighter than 4th (saddle). The IR observations of HS 0810+2554, especially at 2.16 $\mu m$, displayed in Figure 3 of \citealt{2019jones} show the IR flux-ratios are consistent with the standard arrival time of images C being first arriving, etc. 

Although a good fit of $\chi = 1.59$ was found, no model with $\chi \le 1$ was found, even when the number of model parameters exceeded the number of observational constraints. Additionally, Table \ref{tab:res} shows a considerable improvement of $\chi$ values from Mod$_3$ to Mod$_4$. The ability of Mod$_4$ to fit the image positions better than Mod$_3$ is likely owed to the only difference between the models: external shear. As previously mentioned, the influence of nearby massive objects can be represented by an external shear component, i.e., the nearby galaxy group. Because we do explicitly account for the contribution from the nearby group, the large external shear components are likely accounting for missing internal complexity within the mass model itself \citep{1997keeton,2024etherington}. Therefore, the combination of none of our models finding a fit of $\chi\le1$, the best fitting models having large external shear components $\gamma \approx 0.1$, and the position angle of the external shear being somewhat consistent with the position angles of the two EPL's indicate one of two possibilities. First, the galaxy lens could have more mass complexities that we do not include in our mass model, i.e., higher multipoles ($m=5, 6$, etc.), twisting isodensity contours and ellipticity gradients \citep{2022vyvere}, subhalos \citep{2010vegetti,2012vegetti}, edge-on disc \citep{2016hsueh,2017hsueh}, etc.  Second, the perturbations included in our mass models could not be the correct ones and do not approximate the lens. For example, an edge-on disc could be the only higher order perturbation needed to fit the image positions with $\chi\le1$. Or the galaxy could exhibit mass perturbations that are not easily represented by simple parameterization, for example owing to recent or on-going mergers. Alternatively, the perturbations could be produced by different dark matter frameworks, i.e., FDM, that are not easily reproducible outside that framework. However, in preliminary tests (not presented in this paper) we found that mass models containing twistiness, ellipticity gradients, subhalos, and an edge-on disc each by themselves could not provide $\chi<5$. Therefore, the lensing galaxy likely displays slightly more mass complexity than our models offer. With that said, our models still do a good job at reconstructing the image positions. 

An alternative reason for why no mass model with $\chi \le 1$ was found could instead originate from the sources themselves. The eight radio images are modelled as point sources whose image positions were reported by \citealt{2019hartley}. Owed to the high resolution of the European VLBI Network (EVN), the images are localized with astrometric errors two to three times smaller than HST. However, image D2 was resolved into two different sub-components and images A2, B2, and C2 break up into sub-components at higher resolutions. This could indicate that the astrometric errors are under estimated for these images. If the radio positional uncertainties are relaxed to be consistent with those from HST, then our best model Mod$_4$--P$_1^\prime$ would be fit with $\chi \le 1$. We note that the best-fitting model predicted image positions for B1, B2, C1, and C2 are offset from the observed image positions in very narrow azimuthal directions, as shown in Fig. \ref{fig:best}. While these offsets could indicate non-modelled perturbations near these images, they could also be due to errors in image localization. The positional anomalies seen in these images ($\sim$ 10 mas) is on the scale of the separation of sub-components. However, these azimuthal offsets seen in our model predicted images do not seem to correspond to the direction of sub-components observed in \citealt{2019hartley}. In general, we note that issues with the astrometry of one image position could create issues with reconstructing others. 

In an effort to obtain more information about what the lens could look like, we model HS 0810+2554 with the free-form program \texttt{PixeLens} in Appendix \ref{sec:ap1}. \texttt{PixeLens} works by averaging over many mass distributions that perfectly reconstruct the image positions. Thus, the reconstruction we present in Figure \ref{fig:pixelens} shows one possible realization of the lens. Finally, the models presented in Figure \ref{fig:kappas1} and \ref{fig:kappas2} show how different mass models can result in similar $\chi$ values. One of the degeneracies that can be seen in our models is the shape degeneracy. Appendix \ref{sec:ap2} shows and discusses these shape degeneracies.

\section{Conclusion} \label{sec:conc}

The eight radio images of the HS 0810+2554 lensing system have proven to be a challenge for lens reconstructions with simple lens models, thanks to the precise astrometry and close proximity of image pairs in the lens plane. In this paper, we investigate if higher order perturbations to smooth elliptical CDM mass models in the form of Fourier multipoles (i.e., lopsidedness, triangleness, boxiness, and diskiness) are sufficient to explain the positional anomalies of the eight radio images. Because the eight radio images are not source identified and HS 0810+2554 has no measured time-delays, we include all eight possible unique 2x4 configurations (or pairings; P$_{1-8}$) for two arrival time sequences in our analysis (see Table \ref{tab:combs} and Figure \ref{fig:combs}). With four mass models of increasing complexity (Table \ref{tab:models}) and 16 unique image configurations, we find 64 best-fit models of HS 0810+2554 (Table \ref{tab:res}). To the best of our knowledge, no other strong lensing analysis has modelled a single lens as extensively.

Of our 64 best-fit models, we achieve a fit of $\chi=1.59$ with our most complex model Mod$_4$ and image configuration P$_1^\prime$. The image configuration corresponds to the best-fit image pairing found in \citealt{2019hartley} but in the reverse arrival sequence, i.e., images D arrive first, etc. We note that the reverse time ordering is unlikely, but is conceivable if the center of mass of the lens is sufficiently offset from the center of light ($\sim\!20-30$ mas). Time-delay measurements of the HS 0810+2554 system---at any wavelength---would resolve this degeneracy and could be done by programs such as COSMOGRAIL \citep{2005Eigenbrod}, and TDCOSMO \citep{2025dux}, if the source quasar is sufficiently variable. With the morphological arrival time sequence we find a slightly worse, but still adequate, fit of $\chi=2.95$ with image configuration P$_3$, calling into question the true nature of the image pairings. 
Future observations of HS 0810+2554 would be very helpful in determining the image pairings of the radio images and their arrival sequence.

It is intriguing that a dark matter particle with a mass of 10$^{-22}$ eV can, in principle, create perturbations on scales necessary to perfectly reconstruct the eight radio images of HS 0810+2554 \citep{2023amruth}. However, as we have shown in this paper, angular mass complexity in the form of multipoles---and therefore CDM---are adequate in explaining the image positional anomalies in HS 0810+2554, further demonstrating the necessity of their inclusion in future strong lensing analyses.

\section*{Acknowledgments}

The authors would like to thank the anonymous referee for their useful suggestions. The authors acknowledge the Minnesota Supercomputing Institute (MSI) at the University of Minnesota for providing resources that contributed to the research results reported within this paper. URL: http://www.msi.umn.edu. JMJ acknowledges the School of Physics and Astronomy, University of Minnesota, for partially supporting this work through the Edward P. Ney graduate student fellowship. We would like to thank Daniel Gilman for a useful discussion and the suggestion of a baryonic disk. Additionally, we would like to thank Alfred Amruth, Tom Broadhurst, and Jeremy Lim for helpful comments. JMJ would like to thank Derek Perera for discussing at length the work presented here, for attempting to model HS 0810+2554 with \texttt{GRALE}, and overall patience with their shared MSI time. Finally, JMJ would like to thank Io Miller for their editorial oversight and lensing expertise.

\section*{Data Availability}

Data generated from this article will be shared upon reasonable request to the corresponding author.



\bibliographystyle{mnras}
\bibliography{bib} 



\appendix

\section{Quasar Quad} \label{sec:qq}

\begin{figure*}
    \centering
    \includegraphics[width=0.98\linewidth]{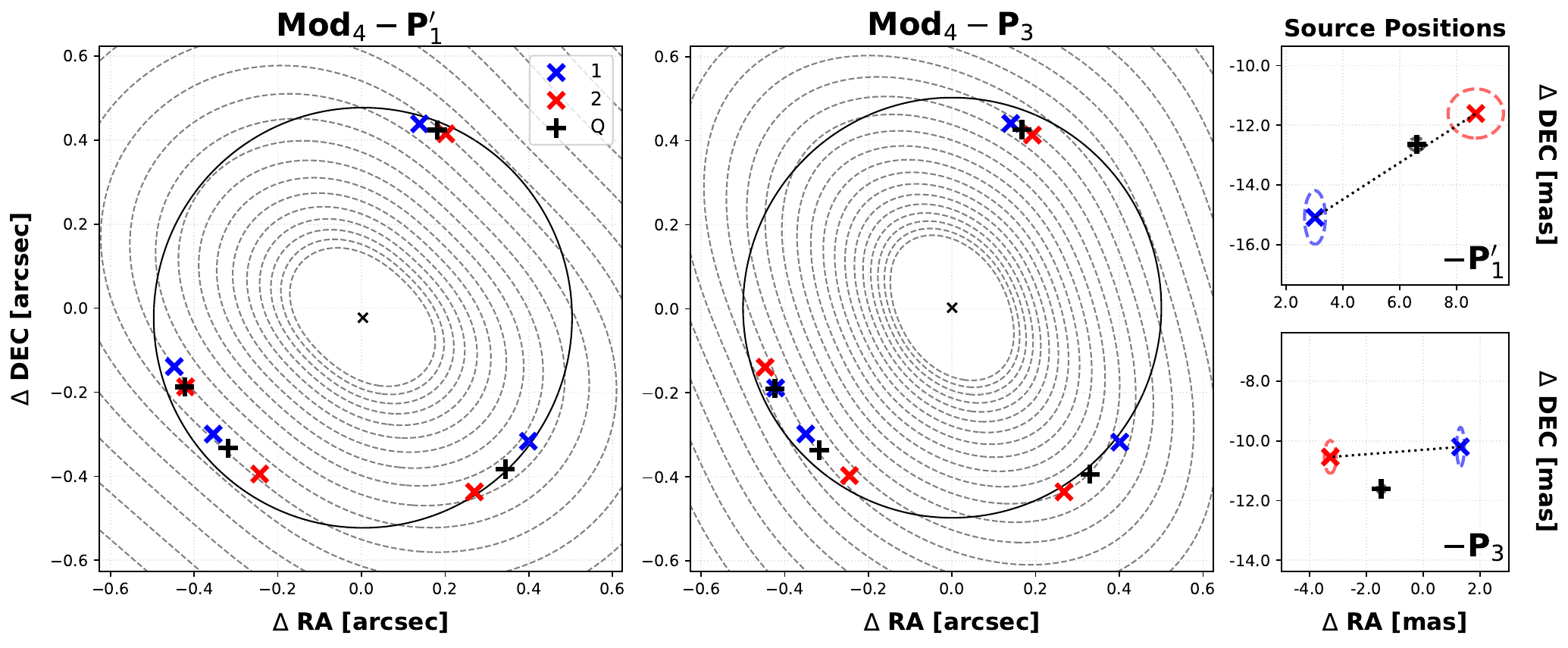}
    \caption{The best-fit quasar source and resulting model image positions (Q; black `+') for models Mod$_4$-P$_1^\prime$ (left) and Mod$_4$-P$_3$ (middle). The image positions are overlaid the mass models $\kappa$ maps (dotted light gray). The gray circle has a radius of 0.5 arcsec and is centered on the galaxy center (black `X'). Similar to Figure \ref{fig:hs0810}, the circle is included to help guide the eye. The two smaller figures on the right display the locations of the three sources for each model (top, Mod$_4$-P$_1^\prime$; bottom, Mod$_4$-P$_3$) with $3\sigma$ error ellipses (the QSO error ellipses are beneath the `+' marker). A black dotted line connecting the two radio sources is included to show the linearity of the three sources.}
    \label{fig:qq}
\end{figure*}

The image positions of the optical quad were not utilized in mass model fitting due to the absence of the lensing galaxy in radio and registration error between optical and radio image positions. Even though the mass models presented in Section \ref{sec:results} were constrained only by the image positions of the two radio quads, one can determine how well the mass model reconstructs the morphology of the optical quad, and thus the entire three quad system. To achieve this the underlying mass model parameters are frozen, leaving only the optical source position $(x_{sQ}, y_{sQ})$ to be optimized. We determine the best fitting quasar source for our two best fitting models Mod$_4$-P$_1^\prime$ and Mod$_4$-P$_3$.

The optical image positions reported in Table \ref{tab:positions} accurately represent the morphology of the image positions (i.e., the distance from one optical image to another), but may not represent the accurate locations of the images with respect to the radio images. That is, the location of the optical images with respect to the radio images in the lens plane can be represented by $(x_i +\Delta x, y_i +\Delta y)$, where $\Delta x$ and $\Delta y$ represent a translation, or systematic shift between the optical and radio coordinate systems, and $(x_i,y_i)$ are the optical image positions in Table \ref{tab:positions}. If Figure \ref{fig:hs0810} depicts the true morphology of the three quad system, then $(\Delta x, \Delta y)$ would equal $(0,0)$. However, this is extremely unlikely and $(\Delta x, \Delta y)$ are likely of the order $\pm 10-20$ mas.

To solve for the quasar source position the `Nelder-Mead' code (see Section \ref{sec:ours}) is initialized to explore the region between the two radio sources in the source plane. After the model images for a given trial source are acquired, the code optimizes for the systematic shift $(\Delta x, \Delta y)$. Then, the fitness of the model images (Eq. \ref{eq:chi}) is determined with the translated observed image positions. The code then draws a new source and repeats. After enough trials, the program attains the best fitting quasar source position and its corresponding shift of the observed image positions in the lens plane. The initial simplex for each model, and thus the initial search region, is generated from the model's two radio source positions.

The best fitting quasar source for our two best fitting mass models Mod$_4$-P$_1^\prime$ and Mod$_4$-P$_3$ are shown in Figure \ref{fig:qq}. In milliarcseconds, the source positions for the two models are $(6.61, -12.64)$ and $(-1.47, -11.61)$, respectively, and the necessary systematic shifts of the observed image positions are $(29.09, -16.53)$ and $(22.30, -20.12)$, respectively. The mass models were not perfect in reconstructing the morphology of the optical quad. The $\chi$ value of Mod$_4$-P$_1^\prime$ increased from $\chi=1.59$ ($N_{\textup{ims}}=8$) to $\chi=3.60$ ($N_{\textup{ims}}=12$) and Mod$_4$-P$_3$ increased from $\chi=2.95$ to $\chi=4.27$. Interestingly, both models differ from Figure \ref{fig:hs0810} by having the optical image AQ overlaid the bottom radio image in A instead of the top.

The two smaller figures in Figure \ref{fig:qq} display the locations of the three source positions for both mass models. If the two radio sources are indeed radio jets stemming from the same central RQQ, then the quasar source would likely fall somewhere between the two radio sources (along the black dotted line). For Mod$_4$-P$_1^\prime$, the quasar source does fall along the line connecting the two radio sources, slightly skewed towards the second radio source (similar to Figure 6 of \citealt{2019hartley}). The quasar source for Mod$_4$-P$_3$ falls about 1 mas vertically below the line connecting the two radio sources, and again is slightly skewed towards the second radio source. The deviation from linearity in P$_3$ could be an indication of jet bending from the intergalactic medium, which appears to become more common at higher redshift \citep{2022morris}.

\section{PixeLens} \label{sec:ap1}

\begin{figure}
    \centering
    \includegraphics[width=0.9\linewidth]{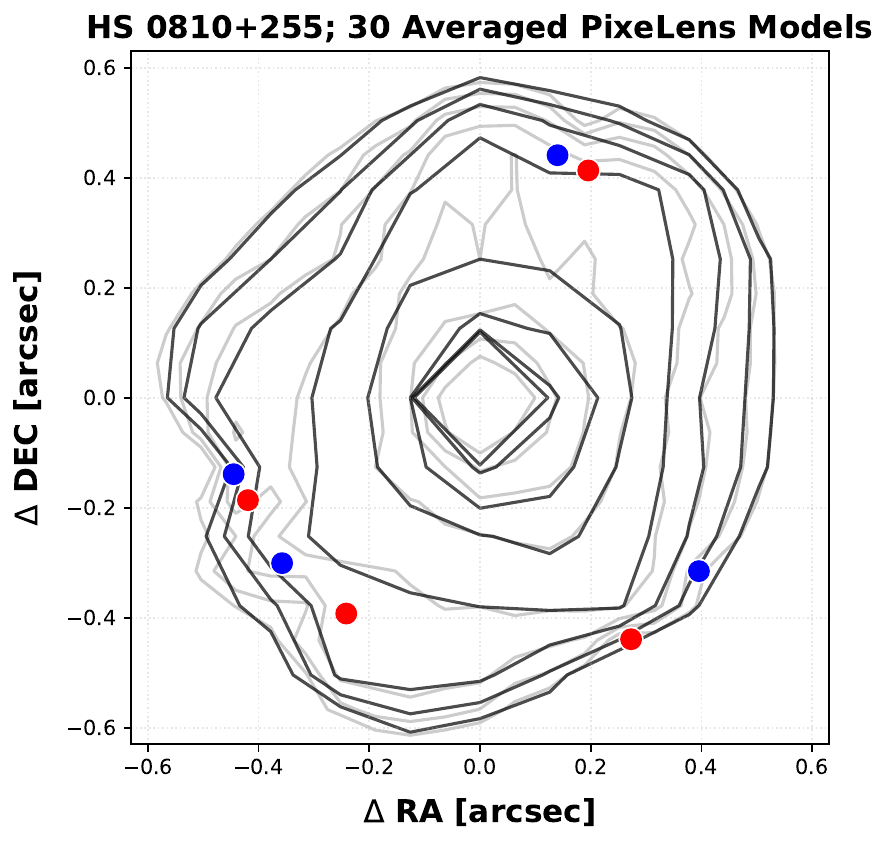}
    \caption{The average convergence of 30 PixeLens models fit to the P$_1$ image configuration. Light gray shows the contours of the 21 $\times$ 21 pixel raw data, and black shows the contours of the 11 $\times$ 11 pixel spline-fit data.}
    \label{fig:pixelens}
\end{figure}

The models presented in Section \ref{sec:results} reproduce the eight radio images well, but no models find $\chi\leq1$. To glean information about what additional mass complexities could exist within the lens, we include a model of HS 0810+2554 created by the free-form galaxy modelling program \texttt{PixeLens} \citep[Figure \ref{fig:pixelens},][]{2004saha}. The free-form nature of the program should allow for perturbations that are not biased by parameterized mass models.

\texttt{PixeLens} works by finding many mass distributions, each of which reconstructs image positions exactly. The final mass map is an ensemble-average of these individual solutions. Being a linear combination of exact solutions, it also perfectly fits the images. As alluded to in the name, the reconstructed mass distribution from \texttt{PixeLens} is coarsely pixelated. To model the eight radio images, we specified the size of the modelling window to be 21 $\times$ 21 pixels with a pixel size of 0.06 arcseconds.

\texttt{PixeLens} depends only on the observed image positions in arrival time order, the redshifts of the lens and source, and a given cosmology.  
Model requirements are implemented such that the resulting mass distribution is reasonably smooth. The $\kappa$ value for any given pixel, besides the center, must be less than or equal to two times the $\kappa$ average of its 8 neighboring pixels, and the density slope must point away from the center to within a specified tolerance angle. See Section 2.2 of \citealt{2004saha} for a detailed discussion of the model. 

Figure \ref{fig:pixelens} was created by generating 30 mass models and randomly shifting the origin of the eight radio image positions within a small window to smear out the inherent pixelation of the program. These offsets are randomized between  $\pm 0.03$ arcseconds (or the width of a pixel). The 30 mass density profiles are then averaged.
We generate only one realization with our analysis, using the P$_1$ image configuration. Because \texttt{PixeLens} generates perfectly fitting mass models, we use it to find one ensemble-averaged model that could represent the lensing galaxy. It is interesting to note that the map in Figure~\ref{fig:pixelens} does not bear close resemblance to those in Figure~\ref{fig:kappas1} for P$_1$. This further suggests that the space of possible solutions is wide, and contains many degenerate solutions, related mostly by shape degeneracies.

\section{Shape Degeneracies} \label{sec:ap2}

\begin{figure*}
    \centering
    \includegraphics[width=\linewidth]{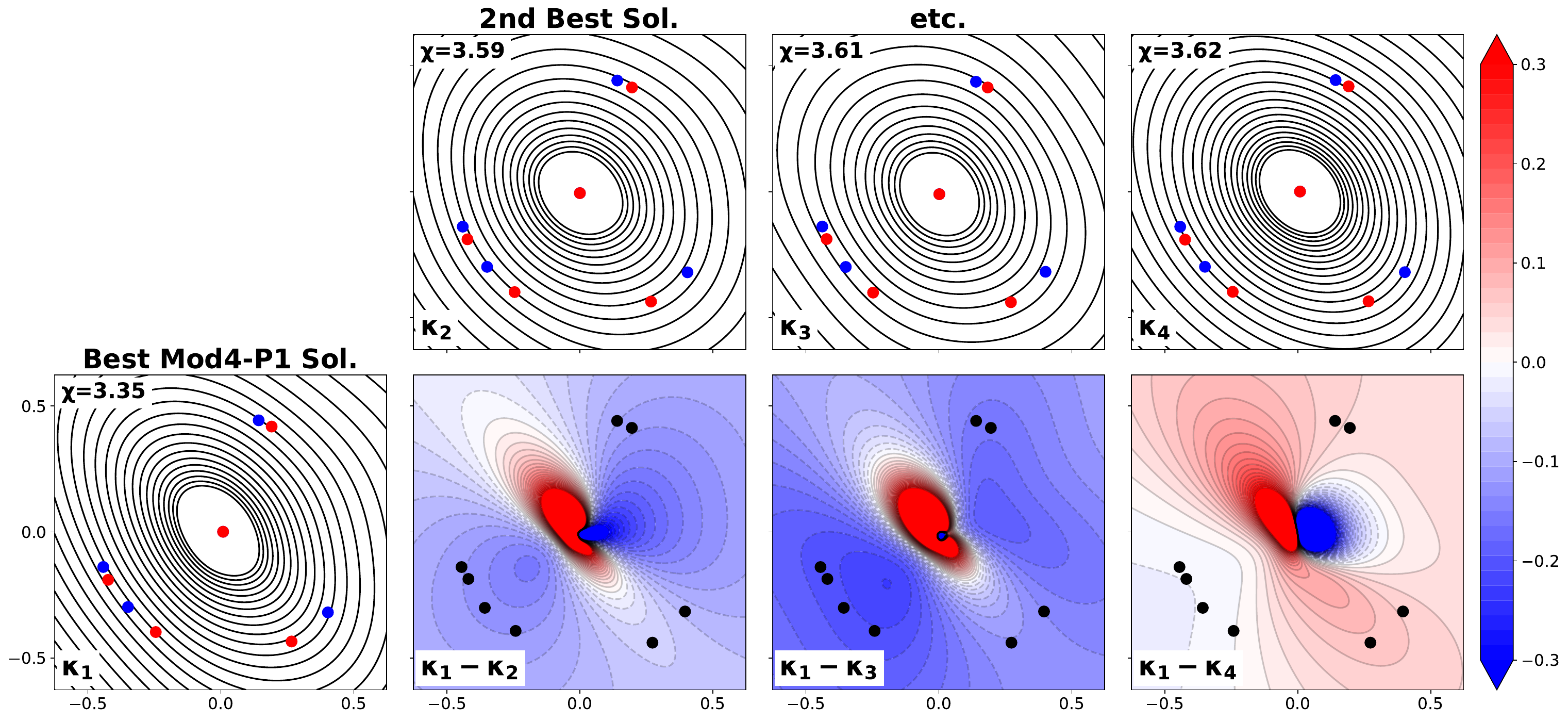}
    \caption{Subtracted convergence, or $\kappa$, maps of the four best fitting Mod$_4$--P$_1$ models, displaying both shape and, a very approximate mass sheet degeneracy. Bottom left shows the $\kappa$ map of the best fitting Mod$_4$--P$_1$ model with $\chi=3.35$, labeled $\kappa_1$. The top row displays the $\kappa$ maps of the three next best fitting Mod$_4$--P$_1$ mass models ($\kappa_{2-4}$) with $\chi = 3.59, 3.61, $and $3.62$, respectively. The bottom row displays the subtraction of the $\kappa$ maps $\kappa_1$ and $\kappa_\#$, where $\kappa_\#$ corresponds to the kappa map above the given figure. Positive subtracted $\kappa$, i.e., $\kappa_1 > \kappa_\#$, is shown in red and solid contours, whereas negative subtracted $\kappa$ is shown in blue and dashed contours. Areas where the two maps are equal are shown in white. The eight images are changed to black in the $\kappa$ subtracted figures to avoid confusion with the contours. The model parameters for these solutions are tabulated in Table \ref{tab:params}. }
    \label{fig:subtraction}
\end{figure*}

It is widely recognized that lensing degeneracies affect recovered mass distribution of the deflectors. Degenerate models are those that reconstruct the properties of observed lens images equally well, but have different mass maps. The best known degeneracy is the mass sheet, or steepness degeneracy \citep{2000saha}, originally identified by \cite{1985falco}. The mass sheet degeneracy is often analyzed in the context of measuring the Hubble constant because it changes the slope of the lens' density profile, but does not affect the mass {\it distribution} in the lens plane. A lesser known degeneracy, but one that does affects the mass distribution, is the monopole degeneracy. This degeneracy allows specific types of redistribution of mass between observed images (but not at their location), without affecting image properties \citep{2012liesenborgs,2024liesenborgs}. 

Originally discussed in \cite{2006saha}, an even broader class of degeneracies that affects the recovered mass distribution are shape degeneracies. This appendix presents examples of these on a galaxy scale. To see a discussion of shape degeneracies in cluster-scale lenses, see \cite{2025perera}. 

To illustrate the shape degeneracies in our models, we compare the $\kappa$ maps for one set of our reconstructed mass maps, those for Mod$_4$--P$_1$, by subtracting the two maps. We note that other models would have produced qualitatively similar illustrations. Figure~\ref{fig:subtraction} shows these $\kappa$ subtractions. All four mass maps, $\kappa_1$ through $\kappa_4$ (panels with black isodensity contour lines) have very similar $\chi$ values, and are therefore degenerate. The differences between these mass maps, i.e., $\kappa_i-\kappa_j$, display the shape degeneracies and are shown in the lower three panels using the red-blue color schemes.

The shapes of these $\kappa$ subtractions are not easily quantifiable. Importantly, the $\kappa_i-\kappa_j$ values are not zero at the locations of the images, meaning that the monopole degeneracy does not play a large role. Additionally, the degeneracies are not of the pure mass sheet type. A very approximate version of the mass sheet degeneracy can be discerned in the two middle lower panels. In both $\kappa_1-\kappa_2$ and $\kappa_1-\kappa_3$, the $\kappa_1$ map has higher amplitude near the center, and lower further out, which means that $\kappa_1$ density profile is steeper than either $\kappa_2$, or $\kappa_3$. Because the degeneracies are not of the pure mass sheet or monopole type, the  differences are best described as {\it shape} degeneracies. 

\section{Model Parameters} \label{sec:params}

The mass model parameters for the best-fit models shown in Figures \ref{fig:kappas1} and \ref{fig:kappas2} are tabulated in Table \ref{tab:params}.

\begin{landscape}
\begin{table}
    \centering
    \begin{tabular}{cccccccccccccccccccc} \hline
       $\chi$ & P$_\#$ & ($x_{s1}$, $y_{s1}$, $x_{s2}$, $y_{s2}$) & ($x_g$, $y_g$) & $t_1$ & $b_1$ & $q_1$ & $\varphi_1$ & $t_2$ & $b_2$ & $q_2$ & $\varphi_2$ & $a_1$ & $\theta_1$ & $a_3$ & $\theta_3$ & $a_4$ & $\theta_4$ & $\gamma$ & $\theta_\gamma$ \\ 
        &  & [mas] & [mas] &  & [mas] &  & [deg] &  & [mas] &  & [deg] &  & [deg] &  & [deg] &  & [deg] &  & [deg] \\ \hline
		\textbf{4.2} & P$_1$ & (-6.3, -7.4, -11.5, -10.1) & (0.5, 3.8) & 0.61 & 151.5 & 0.64 & -126.0 & 1.18 & 173.8 & 0.51 & -47.8 & 1.6e-2 & -168.1 & 2.0e-3 & -72.5 & 1.1e-2 & -87.8 \\ 
		4.78 & P$_1$ & (-22.9, -26.5, -28.0, -28.9) & (-4.2, -13.8) & 0.60 & 114.4 & 0.70 & -109.6 & 1.06 & 239.5 & 0.82 & -29.9 & 4.7e-2 & -160.7 & 4.0e-3 & -52.8 & 9.8e-4 & -15.3 \\ 
		4.79 & P$_1$ & (-23.8, -9.4, -29.5, -12.4) & (-2.8, 2.6) & 0.66 & 143.1 & 0.72 & -123.2 & 1.40 & 237.5 & 0.66 & -57.2 & 4.3e-2 & 178.0 & 5.4e-3 & -57.3 & 2.4e-3 & 177.4 \\ 
		4.82 & P$_1$ & (-7.8, -12.7, -13.9, -16.2) & (3.6, 4.3) & 0.61 & 138.5 & 0.71 & -128.2 & 1.24 & 214.2 & 0.64 & -61.5 & 2.3e-2 & -170.3 & 3.7e-3 & 51.0 & 1.0e-3 & -83.5 \\ \hline
		\textbf{3.79} & P$_3$ & (-16.8, -15.4, -23.2, -16.5) & (-4.5, -4.0) & 0.64 & 180.4 & 0.74 & -129.9 & 1.40 & 188.2 & 0.53 & -58.2 & 3.3e-2 & -169.7 & 4.5e-3 & -58.3 & 9.2e-3 & -106.5 \\ 
		3.84 & P$_3$ & (-12.1, -2.5, -16.6, -2.8) & (-0.0, 5.2) & 0.60 & 99.7 & 0.61 & -133.9 & 1.08 & 217.7 & 0.60 & -58.2 & 2.7e-2 & 164.2 & 3.8e-3 & -66.1 & 5.8e-3 & 76.9 \\ 
		3.87 & P$_3$ & (-2.1, -13.9, -10.2, -15.6) & (1.4, -2.4) & 0.67 & 181.7 & 0.72 & -131.0 & 1.37 & 185.7 & 0.52 & -59.2 & 1.8e-2 & -164.4 & 6.1e-3 & -56.8 & 1.3e-2 & 163.0 \\ 
		3.88 & P$_3$ & (-15.8, -14.7, -22.0, -15.4) & (-2.2, -1.3) & 0.60 & 107.6 & 0.67 & -128.0 & 1.25 & 241.6 & 0.66 & -57.0 & 3.3e-2 & -173.3 & 4.2e-3 & -64.2 & 4.4e-3 & -110.1 \\ \hline
		\textbf{3.35} & P$_1$ & (-21.2, -14.2, -24.6, -16.6) & (-11.5, 4.0) & 1.24 & 144.8 & 0.70 & -84.5 & 1.26 & 285.2 & 0.57 & -50.2 & 2.3e-2 & -143.0 & 4.8e-3 & -99.5 & 3.4e-3 & -157.9 & 0.12 & -46.7 \\ 
		3.59 & P$_1$ & (-12.7, -16.9, -15.9, -19.0) & (0.0, -5.3) & 0.93 & 214.6 & 0.77 & -81.9 & 1.24 & 214.6 & 0.67 & -35.3 & 2.6e-2 & -161.6 & 1.7e-3 & -87.3 & 5.0e-3 & -58.8 & 0.07 & 138.3 \\ 
		3.61 & P$_1$ & (-0.1, -12.9, -3.1, -14.4) & (1.7, -9.5) & 0.60 & 46.6 & 0.54 & -48.8 & 1.09 & 321.2 & 0.85 & -66.4 & 5.6e-3 & -164.6 & 2.1e-3 & -58.9 & 4.1e-3 & 77.3 & 0.10 & -45.6 \\ 
		3.62 & P$_1$ & (7.9, -8.5, 3.7, -11.2) & (8.2, 1.0) & 1.27 & 123.5 & 0.79 & -40.4 & 1.35 & 347.8 & 0.73 & 124.4 & 3.6e-7 & -103.5 & 3.2e-3 & -83.6 & 1.7e-3 & -144.0 & 0.08 & -37.7 \\ \hline
		\textbf{2.95} & P$_3$ & (1.3, -10.2, -3.3, -10.5) & (0.4, 2.4) & 1.10 & 75.0 & 0.50 & -48.5 & 1.36 & 352.5 & 0.72 & -62.9 & 5.1e-7 & 0.5 & 5.3e-3 & 36.4 & 4.5e-3 & 143.0 & 0.10 & -44.9 \\ 
		3.05 & P$_3$ & (2.4, -6.1, 0.2, -6.6) & (2.0, -1.9) & 1.15 & 143.6 & 0.80 & -129.4 & 1.23 & 282.3 & 0.57 & -55.3 & 3.8e-6 & 86.6 & 1.3e-3 & -91.1 & 6.2e-3 & 170.9 & 0.09 & -48.1 \\ 
		3.16 & P$_3$ & (-0.1, -6.8, -6.9, -7.9) & (-1.6, 7.2) & 0.60 & 77.0 & 0.69 & -138.4 & 1.33 & 244.8 & 0.50 & -62.1 & 3.5e-6 & 10.1 & 5.9e-3 & -70.8 & 1.1e-2 & -112.6 & 0.05 & 130.5 \\ 
		3.17 & P$_3$ & (2.2, -3.2, -2.3, -4.2) & (-0.3, 3.0) & 0.85 & 171.6 & 0.64 & -64.0 & 0.91 & 214.4 & 0.88 & 147.4 & 2.0e-7 & -45.4 & 3.7e-3 & -65.5 & 6.9e-3 & -117.4 & 0.09 & -46.4 \\ \hline\hline
		\textbf{4.59} & P$_1^\prime$ & (-7.8, -9.7, -1.2, -6.6) & (5.5, -24.5) & 0.66 & 91.9 & 0.68 & -135.0 & 1.40 & 257.3 & 0.58 & -48.8 & 2.6e-2 & 155.2 & 3.4e-3 & -7.6 & 1.1e-2 & -90.4 \\ 
		4.88 & P$_1^\prime$ & (-25.6, 0.4, -20.3, 3.2) & (-0.5, -21.2) & 0.60 & 104.9 & 0.75 & -95.0 & 1.32 & 256.0 & 0.70 & -15.1 & 5.6e-2 & 159.5 & 3.2e-3 & -27.2 & 3.8e-10 & -16.4 \\ 
		4.93 & P$_1^\prime$ & (9.3, -19.6, 18.3, -15.4) & (15.3, -38.5) & 0.62 & 119.0 & 0.80 & -118.4 & 1.39 & 240.3 & 0.60 & -33.9 & 1.1e-2 & 173.0 & 3.5e-3 & -127.0 & 8.1e-3 & 1.6 \\ 
		4.98 & P$_1^\prime$ & (-1.5, 26.9, 0.7, 27.9) & (11.7, 2.2) & 0.63 & 63.6 & 0.69 & -116.7 & 1.16 & 305.6 & 0.80 & -42.0 & 5.4e-2 & 118.2 & 1.7e-8 & -18.3 & 4.2e-3 & -74.8 \\ \hline
		\textbf{4.77} & P$_3^\prime$ & (25.0, -0.1, 27.6, 0.3) & (24.4, -12.1) & 0.60 & 106.7 & 0.74 & -105.9 & 1.40 & 263.8 & 0.70 & -26.9 & 1.6e-2 & 79.2 & 7.2e-4 & 5.8 & 5.8e-3 & -70.7 \\ 
		5.16 & P$_3^\prime$ & (-2.6, 8.7, -0.2, 9.0) & (8.9, -8.8) & 0.60 & 156.3 & 0.84 & -124.6 & 1.32 & 217.1 & 0.68 & -49.0 & 3.6e-2 & 127.5 & 8.6e-4 & -118.1 & 5.0e-3 & -84.4 \\ 
		5.24 & P$_3^\prime$ & (20.1, -18.3, 23.5, -18.2) & (20.6, -27.3) & 0.60 & 92.3 & 0.79 & -121.5 & 1.35 & 277.3 & 0.72 & 138.0 & 1.4e-7 & 101.2 & 2.3e-3 & 1.4 & 6.6e-3 & -175.6 \\ 
		5.37 & P$_3^\prime$ & (18.6, 0.6, 21.3, 1.1) & (20.3, -11.9) & 0.89 & 176.1 & 0.73 & -113.5 & 1.40 & 244.5 & 0.66 & -36.0 & 2.0e-2 & 92.0 & 1.7e-3 & 12.8 & 8.1e-3 & -73.9 \\ \hline
		\textbf{1.59} & P$_1^\prime$ & (3.0, -15.1, 8.7, -11.6) & (3.4, -22.7) & 1.01 & 165.5 & 0.63 & -28.0 & 1.05 & 231.2 & 0.66 & -51.2 & 8.5e-3 & -53.7 & 4.0e-3 & -14.6 & 5.8e-3 & -92.0 & 0.12 & -37.7 \\ 
		1.59 & P$_1^\prime$ & (3.0, -15.1, 8.7, -11.6) & (3.4, -22.7) & 1.01 & 165.5 & 0.63 & -28.0 & 1.05 & 231.2 & 0.66 & -51.2 & 8.5e-3 & -53.7 & 4.0e-3 & -14.6 & 5.8e-3 & -92.0 & 0.12 & -37.7 \\ 
		1.75 & P$_1^\prime$ & (-8.7, -21.3, 2.6, -14.5) & (-1.6, -32.8) & 0.97 & 150.3 & 0.57 & -29.8 & 1.35 & 247.4 & 0.58 & -41.8 & 1.9e-2 & -109.9 & 9.0e-3 & -8.0 & 1.2e-2 & -88.0 & 0.12 & -34.7 \\ 
		1.82 & P$_1^\prime$ & (-4.8, 24.2, 6.4, 30.4) & (-3.4, -3.4) & 1.32 & 133.5 & 0.63 & -65.8 & 1.33 & 282.7 & 0.52 & -45.8 & 3.6e-2 & 78.1 & 4.2e-3 & -133.2 & 1.2e-2 & -86.8 & 0.12 & -48.4 \\ \hline
		\textbf{2.03} & P$_3^\prime$ & (-14.3, -10.8, -8.2, -10.4) & (-6.3, -24.0) & 0.92 & 176.8 & 0.63 & -57.2 & 1.15 & 201.9 & 0.54 & -27.3 & 1.2e-2 & -145.0 & 6.8e-3 & -18.0 & 1.1e-2 & -86.6 & 0.13 & -39.7 \\ 
		2.15 & P$_3^\prime$ & (-12.5, -3.5, -4.7, -1.8) & (-5.3, -12.4) & 0.61 & 72.7 & 0.87 & -55.7 & 1.16 & 253.6 & 0.54 & -46.2 & 4.6e-3 & 111.4 & 6.3e-3 & -4.1 & 1.3e-2 & 8.7 & 0.12 & -44.5 \\ 
		2.28 & P$_3^\prime$ & (-6.7, 1.6, 0.0, 2.6) & (0.7, -15.5) & 0.97 & 147.4 & 0.56 & 142.6 & 1.28 & 265.9 & 0.70 & -41.3 & 1.2e-2 & 126.6 & 4.6e-3 & -10.6 & 1.2e-2 & -172.9 & 0.12 & -36.2 \\ 
		2.35 & P$_3^\prime$ & (-7.4, 13.6, 0.7, 13.9) & (0.5, -18.9) & 0.99 & 112.1 & 0.67 & -42.1 & 0.99 & 271.4 & 0.62 & -28.6 & 2.4e-2 & 113.5 & 6.4e-3 & -21.7 & 1.4e-2 & 4.4 & 0.12 & -30.3 \\  \hline
    \end{tabular}
    \caption{The model parameters for the models shown in Figure \ref{fig:kappas1} (top; standard time ordering) and Figure \ref{fig:kappas2} (bottom; reverse time ordering). The models are Mod$_{3-4}$ for image configurations P$_{1,3}$ and P$_{1,3}^\prime$. Models that represent the best-fit solution for its model and image configuration, as tabulated in Table \ref{tab:res}, have their $\chi$ value in \textbf{bold}. P$_\#$ corresponds to the image configuration of the model; $x_{s1}$, $y_{s1}$, $x_{s2}$, and $y_{s2}$ are the two source positions; $x_{g}$ and $y_{g}$ the center of the galaxy (or equivalently the EPL profiles); $t, b, q,$ and $\varphi$ are the slope, normalization, axis-ratio, and position angle of the two EPL profiles; $a_m$ and $\theta_m$ are the normalization and position angle for the given multipole $m$; and, $\gamma$ and $\theta_\gamma$ are the normalization and position angle of the external shear. Models with no reported $\gamma$ and $\theta_\gamma$ are from Mod$_3$. All models have an additional constant external shear of $\gamma$ = 0.026 and $\theta_{\gamma} = 91.5 \degree$ to account for the nearby galaxy group. The source and galaxy positions are centered on the CASTLES galaxy center with a (+6 mas, +4 mas) offset, as discussed in Section \ref{sec:lensobs}.}
    \label{tab:params}
\end{table}
\end{landscape}

\bsp	
\label{lastpage}
\end{document}